\input harvmac

\def \const {{\rm const}}

\def \ra {\rangle}
 
\def \ov {\over}

\def \ep {\epsilon}
\def \k {\kappa}

\def \L {{\cal L}}

\def \J {{\cal J}}

\def \ss {{\cal S}}

\def \pa { \partial}
\def \a {\alpha}
\def \E {{\cal E}}
\def \b {\beta}
\def \g {\gamma}
\def \G {\Gamma}
\def \d {\delta}

\def \l {\lambda}

\def \m {\mu}
\def \n {\nu}

\def \s {\sigma}

\def \r {\rho}
\def \t {\theta}
\def \ta {\tau}
\def \p {\phi}
\def \P { \Phi}

\def \vp {\varphi}

\def \frac#1#2{{ #1 \over #2}}
\def \lr { \lref}
\def \td {\tilde}

\def \aa {{\a'}}

\def \lr{\lref}

\def \rf {\refs}

\def \ee {{\rm e}}

\def \adss {$AdS_5 \times S^5\ $}
\def \ads {$AdS_5$\ }

\def \ta { \tau}
\def \s { \sigma }

\def \vp {\varphi}

\def \p {\phi}
\def \vt {\theta}
 \def \a { \alpha}
\def \r {\rho}
\def \fourth {{1 \ov 4}}
\def \fo  {{{\textstyle {1 \ov 4}}}}

\def \vr {\varrho}
  \def \td { \tilde }

\def \t {\theta}

\def \del{\partial}
\def \m {\mu }
\def \n {\nu }
\def \ha { { 1 \over 2}}

\def \vr  { \varrho}

\def \g {\gamma}
\def \G {\Gamma}
\def \k {\kappa}
\def \l {\lambda}
\def \L {{\cal L}}

\def \td {\tilde }
\def \b{\beta}

\def \tr {{\rm tr}}
\def \ha {{1 \over 2}}
\def \ep{\epsilon}

\def \ov {\over}
\def \JJ {{\cal J}} \def \Om {\Omega}

\def \z {\zeta}

\def \w {w}

\def \om {\kappa}
\def \w  {\omega}

\def \ww { {\rm w} }

\def \sql {{\sqrt{\l}}\ }

\def \ta {\tau}
\def \tdr {{\td \r}}

\def \D {{\Delta}}
\def \E {{\cal E}}
\def \JJ {{J'}}
\def \ta {\tau}
\def \JJJ {{\cal J}'}
\def \L {\Lambda}
\def \II {{\cal V}}

\lr \ftn{S. Frolov and A.A. Tseytlin, to appear.}

\lr \lip{
A.~V.~Kotikov, L.~N.~Lipatov and V.~N.~Velizhanin,
``Anomalous dimensions of Wilson operators in N = 4 SYM theory,''
Phys.\ Lett.\ B {\bf 557}, 114 (2003)
[hep-ph/0301021].
}

\lref\arut{
G.~Arutyunov, B.~Eden, A.~C.~Petkou and E.~Sokatchev,
``Exceptional non-renormalization properties and OPE analysis of
chiral  four-point functions in N = 4 SYM(4),'' Nucl.\ Phys.\ B
{\bf 620}, 380 (2002) [hep-th/0103230].
}

\lr \dev { H.~J.~de Vega and I.~L.~Egusquiza,
``Planetoid String Solutions in 3 + 1 Axisymmetric Spacetimes,''
Phys.\ Rev.\ D {\bf 54}, 7513 (1996)
[hep-th/9607056].
 }

  \lr\arm{
A.~Armoni, J.~L.~Barbon and A.~C.~Petkou,
``Orbiting strings in AdS black holes and N = 4 SYM at finite
temperature,''
JHEP {\bf 0206}, 058 (2002)
[hep-th/0205280].
``Rotating strings in confining AdS/CFT backgrounds,''
JHEP {\bf 0210}, 069 (2002)
[hep-th/0209224].
}

\lr \oog {  G.~T.~Horowitz and H.~Ooguri,
``Spectrum of large N gauge theory from supergravity,''
Phys.\ Rev.\ Lett.\  {\bf 80}, 4116 (1998)
[hep-th/9802116].
E.~Witten,
``Anti-de Sitter space and holography,''
Adv.\ Theor.\ Math.\ Phys.\  {\bf 2}, 253 (1998)
[hep-th/9802150].
T.~Banks, M.~R.~Douglas, G.~T.~Horowitz and
E.~J.~Martinec,
``AdS dynamics from conformal field theory,''
hep-th/9808016.
}

\lr \KT{
R.~Kallosh and A.~A.~Tseytlin,
``Simplifying superstring action on \adss,''
JHEP {\bf 9810}, 016 (1998)
[hep-th/9808088].
S.~Forste, D.~Ghoshal and S.~Theisen,
``Stringy corrections to the Wilson loop in N = 4 super
 Yang-Mills
theory,''
JHEP {\bf 9908}, 013 (1999)
[hep-th/9903042].
}

\lr \fadd{
L.~D.~Faddeev,
``How Algebraic Bethe Ansatz works for integrable model,''
hep-th/9605187.
}

\lr \bie{
N.~Beisert, C.~Kristjansen and M.~Staudacher,
``The dilatation operator of N = 4 super Yang-Mills theory,''
hep-th/0303060.
}
\lr\dol{
F.~A.~Dolan and H.~Osborn,
``Superconformal symmetry, correlation functions and the operator
 product
expansion,''
Nucl.\ Phys.\ B {\bf 629}, 3 (2002)
[hep-th/0112251].
}

\lr \bmn { D.~Berenstein, J.~Maldacena and H.~Nastase,
``Strings in flat space and pp waves from N = 4 super Yang
 Mills,''
JHEP {\bf 0204}, 013 (2002)
[hep-th/0202021].
}

\lr \gkp {
S.~S.~Gubser, I.~R.~Klebanov and A.~M.~Polyakov,
``A semi-classical limit of the gauge/string correspondence,''
Nucl.\ Phys.\ B {\bf 636}, 99 (2002)
[hep-th/0204051].
}

\lr \pol {A.~M.~Polyakov,
``Gauge fields and space-time,''
Int.\ J.\ Mod.\ Phys.\ A {\bf 17S1}, 119 (2002)
[hep-th/0110196].
}

\lref\mets{
R.~R.~Metsaev,
``Type IIB Green-Schwarz superstring in plane wave Ramond-Ramond
background,''
Nucl.\ Phys.\ B {\bf 625}, 70 (2002)
[hep-th/0112044].
R.~R.~Metsaev and A.~A.~Tseytlin,
``Exactly solvable model of superstring in plane wave
 Ramond-Ramond
background,''
Phys.\ Rev.\ D {\bf 65}, 126004 (2002)
[hep-th/0202109].
}

\lref\MT{
R.~R.~Metsaev and A.~A.~Tseytlin,
``Type IIB superstring action in \adss  background,''
Nucl.\ Phys.\ B {\bf 533}, 109 (1998)
[hep-th/9805028].
}

\lr \dgt {N.~Drukker, D.~J.~Gross and A.~A.~Tseytlin,
``Green-Schwarz string in \adss : Semiclassical partition
function,''
JHEP {\bf 0004}, 021 (2000)
[hep-th/0001204].
}

\lr \tse { A.~A.~Tseytlin,
``On semiclassical approximation and spinning string vertex
operators in $AdS_5 \times  S^5$,''
hep-th/0304139.
}

\lr\sta{
N.~Beisert, C.~Kristjansen, J.~Plefka, G.~W.~Semenoff and
M.~Staudacher,
``BMN correlators and operator mixing in N = 4 super Yang-Mills
theory,''
Nucl.\ Phys.\ B {\bf 650}, 125 (2003)
[hep-th/0208178].
N.~Beisert,
``BMN operators and superconformal symmetry,''
hep-th/0211032.
N.~Beisert, C.~Kristjansen, J.~Plefka and M.~Staudacher,
``BMN gauge theory as a quantum mechanical system,''
Phys.\ Lett.\ B {\bf 558}, 229 (2003)
[hep-th/0212269].
}
\lr \jon{H.F. Jones, ``Groups, representations and physics'',
IOP Publishing, Bristol and Philadelphia (1998). }

\lr \fts{ S.~Frolov and A.~A.~Tseytlin,
``Semiclassical quantization of rotating superstring in
\adss,''
JHEP {\bf 0206}, 007 (2002)
[hep-th/0204226].
}

\lr\rus{ J.~G.~Russo,
``Anomalous dimensions in gauge theories from rotating strings in
\adss,''
JHEP {\bf 0206}, 038 (2002)
[hep-th/0205244].
}

\lr \ogu{
J.~M.~Maldacena and H.~Ooguri,
``Strings in AdS(3) and SL(2,R) WZW model. I,''
J.\ Math.\ Phys.\  {\bf 42}, 2929 (2001)
[hep-th/0001053].
J.~M.~Maldacena, J.~Michelson and A.~Strominger,
``Anti-de Sitter fragmentation,''
JHEP {\bf 9902}, 011 (1999)
[hep-th/9812073].
N.~Seiberg and E.~Witten,
``The D1/D5 system and singular CFT,''
JHEP {\bf 9904}, 017 (1999)
[hep-th/9903224].
}
\lr \miz{
J.~A.~Minahan and K.~Zarembo,
``The Bethe-ansatz for N = 4 super Yang-Mills,''
JHEP {\bf 0303}, 013 (2003)
[hep-th/0212208].
}

\lr \aha { O.~Aharony, S.~S.~Gubser, J.~M.~Maldacena, H.~Ooguri
and Y.~Oz,
``Large N field theories, string theory and gravity,''
Phys.\ Rept.\  {\bf 323}, 183 (2000)
[hep-th/9905111].
}
\lr \papd{
M.~Blau, M.~O'Loughlin, G.~Papadopoulos and A.~A.~Tseytlin,
``Solvable models of strings in homogeneous plane wave
backgrounds,''
hep-th/0304198.
}

\lr \gor{
A.~V.~Belitsky, A.~S.~Gorsky and G.~P.~Korchemsky,
``Gauge / string duality for QCD conformal operators,''
hep-th/0304028.
}
\lr \hok{E.~D'Hoker and D.~Z.~Freedman,
``Supersymmetric gauge theories and the AdS/CFT correspondence,''
hep-th/0201253.
}

\lr \kw {
I.~R.~Klebanov and E.~Witten,
``Superconformal field theory on threebranes at a Calabi-Yau
singularity,''
Nucl.\ Phys.\ B {\bf 536}, 199 (1998)
[hep-th/9807080].
}

\Title{\vbox
{\baselineskip 10pt
{\hbox{  OHSTPY-HEP-T-03-007}
}}}
{\vbox{\vskip -30 true pt
\medskip
\centerline {Multi-spin  string  solutions
  in $AdS_5 \times S^5$ }
\medskip
\vskip4pt }}
\vskip -20 true pt
\centerline{S. Frolov$^{a,}$\footnote{$^*$} {Also at Steklov
Mathematical Institute, Moscow.}
 and
A.A.~Tseytlin$^{a,b,}$\footnote{$^{**}$}
{Also at
Lebedev Physics Institute, Moscow.}
}
\smallskip\smallskip
\centerline{ $^a$ \it  Department of Physics,
 The Ohio State University,
 Columbus, OH 43210, USA}

\centerline{ $^b$ \it  Blackett Laboratory,
 Imperial College,
 London,  SW7 2BZ, U.K.}

\bigskip\bigskip
\centerline {\bf Abstract}
\baselineskip12pt
\noindent
\medskip

Motivated by attempts to extend AdS/CFT duality to non-BPS states
we consider classical closed string solutions with several
angular momenta in different directions of $AdS_5$ and $S^5$. We
find a novel solution describing a circular closed string located
at a  fixed value of $AdS_5$ radius while rotating simultaneously
in two  planes in $AdS_5$ with equal spins $S$. This  solution is
a direct generalization of a two-spin flat-space solution where
the string rotates in two orthogonal planes while always  lying
on a  3-sphere. Similar solution exists  for a string rotating in
$S^5$: it is parametrized by the angular momentum $J$ of the
center of mass  and two equal $SO(6)$  angular momenta
$J_2=J_3=J'$ in the two  rotation planes. The remarkably simple
case is of $J=0$  where the energy depends on $J'$ as
$E=\sqrt{(2J')^2 + \l}\ $\ \ \ ($\l$  is the string tension or `t
Hooft coupling). We discuss interpolation of the $E(J')$  formula
to weak coupling by identifying the gauge theory operator that
should be dual to the corresponding semiclassical string state
and utilizing  existing results for its  perturbative anomalous
dimension. This opens up a possibility of studying  AdS/CFT duality
in this new non-BPS sector. We also investigate  small
fluctuations and stability of these classical solutions and
comment on several generalizations.

\bigskip

\Date{04/03}

\noblackbox
\baselineskip 16pt plus 2pt minus 2pt

\newsec{Introduction}

Generalizing AdS/CFT duality to non-BPS  string mode sector can
 be
guided by semiclassical considerations,  as suggested  in
\rf{\bmn,\gkp}. Identifying classical solitonic solutions  of
 $AdS_5
\times S^5$  sigma model carrying basic global  charges  is
 important
in order to understand the  structure of the  full string theory
spectrum.  In general, the states belonging to representations
 of the
isometry group $SO(2,4) \times SO(6)$ are expected to be
 classified by
6=3+3  charges corresponding  to Cartan subalgebra generators,
$(E,S_1,S_2; \ J_1,J_2,J_3)$.  Here $S_1$ and $S_2$ are the two
spins of the conformal group  (labelling representations of
 $SO(4)$ 
isometry of $S^3 $ subspace)  and $J_i$ are the three angular
momenta of the $S^5$ isometry  group.
One may search for classical string solutions  which have minimal
 energy
for given values of the  5  charges,
$E=E(S_1,S_2, J_1, J_2, J_3)$.\foot{The Casimir operators should
 be
functions of these  charges. In general, string theory ``knows''
about many other conserved charges, being integrable.  Here we
concentrate on most obvious local charges.} The importance of
 such
solutions (in contrast to various
other oscillating or pulsating   solutions) is that having
non-zero global charges simplifies identification of the
 corresponding  dual CFT operators.

Particular  classical string solutions with  special
 combinations of
these charges were discussed in the past.  Point-like string
 solution
(geodesic) lying in $AdS_5$ does not carry intrinsic  spin.
 Geodesic
running in $S^5$ can carry  only {\it one} component  of momentum
  in
$S^5$ (e.g., $J=J_1$), and expansion of \adss string theory  near
 such
geodesic was studied in \bmn. Extended string solution describing
folded closed string rotating in a plane in $AdS_5$   carries
 single
spin, e.g., $S=S_1$ \rf{\dev,\gkp}. One can boost  the center of
 mass
of the string rotating in $AdS_5$ along a circle of $S^5$
 getting a
solution with two charges $(S,J)$ \fts. Alternatively, one can
construct a  solution describing folded string rotating about a
 pole
of $S^5$ \gkp; while  it carries again only one component (say,
$\JJ=J_2$)  of the $SO(6)$ spin it is not equivalent to a
 point-like
orbiting solution.\foot{The corresponding  string vertex
 operators
\tse\ as well as dual gauge theory operators  should be
 different,
with the ``point-like'' (BPS) one having minimal energy
 (dimension)
for a given value of the angular momentum (R-charge).} An
interpolating solution with the three charges $(S,J,\JJ)$ was
constructed in \rus. One may think  that while in general there
should certainly be  extended string solutions  with more spins
 in
either or both $AdS_5$ and $S^5$ spaces,  they may be difficult
 to
construct explicitly, and also their AdS/CFT interpretation may
 be
unclear. Here we would like to point out that such more general
solutions are actually easy to find in the special case when the
two spins $S_1,S_2$ in $AdS_5$  or the  two of the three angular
momenta in $S^5$ (e.g., $J_2$ and $ J_3$) are {\it equal}.  The
analytic form of the solution with $S_1=S_2\equiv S$  turns out
to be much simpler than in the single spin case of
 \rf{\dev,\gkp}:
it is a direct generalization of the  flat-space solution
 describing
circular  string rotating simultaneously in the two orthogonal
 spatial
planes with equal angular  momenta. Such ``self-dual'' string
 always
lies on an $S^3$ surface in Minkowski space,  and thus it can be
easily ``embedded'' into $AdS_5$ space  using  global (or
 covering)
coordinates. The string  is positioned  at a fixed value of the
$AdS_5$ radius $\r =\r_0$, being stabilized by rotation. Similar
string  solution exists for a more general class of 5-d metrics
$ds^2 = - g(\r) dt^2 +  d\r^2 + h(\r) d\Omega_3$, but not in
 $AdS_3$
or $AdS_4$: to stabilize the circular string at a fixed  value of
$\r$ one  needs at least {\it two} equal  spin
 components.\foot{Note
that   a  similar (but non-rotating)  ``winding'' string
 configuration
near the boundary of $AdS_3$  is stabilized by the $B_{mn}$ flux
\ogu.}

We will  show that this  stationary solution is  stable under
 small
perturbations if the spins  $S_1=S_2=S$ are smaller than  a
 critical
value.   The energy $E$  is an  algebraic function of $S$. For a
small-radius string  having small  $ \ss \equiv {S\ov \sql} \ll
 1$,
i.e. located  close to the center of $AdS_5$, one finds the usual
Regge trajectory relation, $E=( \sql  4S )^{1/2} + ...$.
For large string   located  close to the boundary of $AdS_5$ we
 get
$E = 2S + c_1  (\l S)^{1/3} + ...$ \ ($ \ss \gg1$).
\foot{Interestingly, the power of $\l$ that multiplies a power of
$S$ in the semiclassical correction to $E-2S$ is the same as the
 power
of $S$ only when it is  equal to  1/3. One may   speculate  that
 a
weak-coupling interpolation of this formula is then
$E-2S= (1 + c \l S)^{1/3} $.}
The leading $E=2S$ behaviour is the same as for the single-spin
 folded
string solution \dev, but the subleading correction here is
proportional to $S^{1/3}$ instead of $\ln S$ in \gkp.

Modulo the problem of instability of the two-spin solution  at
 large
$S \gg \sql$, it is natural to conjecture (see section 5) that
 the
Euclidean gauge theory operator in $R^4$ that should be dual to
 this
two-spin string state should  have  the form $\tr[ \Phi_M (D_1 +
 i
D_2)^{S} (D_3 + i D_4)^{S} \Phi_M]+...$, 
 where $\P_M$ are  SYM
 theory
scalars. It would be  interesting to find how  its perturbative
anomalous  dimension  $\D$ depends on large $S$.
An interpolation formula suggested by the semiclassical analysis
 is $
\Delta (S) = S +  f(\l) S^{1/3} + ...$, $S \gg 1 $, where
$f(\l)_{\l \ll 1} = a_1 \l + a_2 \l^2 + ...$, and
$f(\l)_{\l \gg 1} =  \l^{1/3} [c_1   + {c_2 \ov \sql}  + ...]$,
As we shall see   below (following a similar discussion in the
single-spin case \fts), if one formally ignores the instability,
the 1-loop string correction to the energy of the semiclassical
 2-spin
solution does scale  with spin as  $S^{1/3}$. However, it seems
implausible  that a perturbative anomalous dimension may depend
 on the
spin as  a fractional power. We shall comment on this further in
section 3.

 \bigskip

The AdS/CFT  correspondence seems easier to establish in a
 different
sector corresponding to the string solution carrying  $SO(6)$
 spins,
i.e. rotating in $S^5$.  Indeed, there  exists a similar circular
string solution with two equal angular momenta   in $S^5$: the
rotating string moves on $S^3$ within $S^5$, with the   radius
 of the
string or of  $S^3$ being related to the value of $\JJ$. In
 addition,
the center of mass of the string may be rotating along another
circle of $S^5$,  leading to  a particular string   solution
 with
all the three  $S^5$ charges  being non-zero $(J_3=J,\  J_1= J_2
 =\JJ)$.
In the most transparent case of $J=0$ when the string has
 maximal
size  (so that $2J' \geq \sql$) the energy  turns out to depend
 on
$\JJ$ in a remarkably simple way: $E= \sqrt{(2\JJ)^2 + \l} $.
While the solution with $J=0$ turns out to be unstable, 
there is always a non-trivial region of stability  when 
$J\not=0$, and there are stable solutions with both $J$ and $J'$
being large compared to $\sql$ \ftn.

We suggest that the  corresponding dual CFT operator (having
 minimal
canonical dimension for given  values of R-charges $J$ and
$J_2=J_3=J'$)   should be $\tr [ (\Phi_1 + i \Phi_2)^J (\Phi_3 +
 i
\Phi_4)^{\JJ} (\Phi_5 + i\Phi_6)^{\JJ}]+ ...,  $ where dots stand
 for
appropriate permutations of factors. For $J=0$ the above
 semiclassical
formula $E(J')$ suggests  that  for large $J'$  the anomalous
 dimension
of such operator  should  be $\D= 2J' + {f(\l)\ov 4 J'}  + ... $.
We conjecture that $f(\l)$ should start with  a $ \l$-term   both at
strong and weak  coupling, and  we propose to 
check this against
known \rf{\sta,\miz} perturbative results. Another interesting
 direction
 is to
consider the limit  $J\gg J'$  and  relate the resulting
 expression
for the energy  to the dimensions of operators in the  sector
 studied
in \bmn.

\bigskip

The paper is organized as follows. In section 2  we first
 describe
the two-spin closed string  solution in flat space  and then
generalize it to $AdS_5$ and $S^5$  spaces viewed  as
 hypersurfaces
in $R^{2,4}$ and $R^6$ respectively. In section 3 we rederive
 the
$AdS_5$ solution  and study it  in more detail using explicitly
 the
standard set of global coordinates in  $AdS_5$ space.  In
 particular,
we obtain  the action for small  fluctuations near this solution
 and
find that the solution is  stable only if  the spin  is bounded
 from
above, $  S \leq a \sql $, where $a$  is of order 1.

In section 4 we present  a similar analysis of the $S^5$
 solution. We
also note that there exists a ``combined'' solution  where string
rotates in both $AdS_5$ and $S^5$ and thus carries  5 charges:
$S_1=S_2, J=J_1$ and $J_2=J_3$.  We observe that there are  two
branches of the solution with  $S=0$ and $J=0$ with different
dependence of the energy on $J'$:  one  for $ J' \leq \ha \sql$
 and
another for $ J' \geq \ha \sql$.  
The solution with   $ J' \leq \ha \sql$
is found to be  stable for $J' \leq \sql { 3 \ov 8}$, 
while  the solution with large $J'$ 
and $J=0$ turns out to be unstable. 
Given that there are more general stable solutions 
with both $J'$ and $J$ being large \ftn, we shall assume that the
instability may  not preclude one from using the
remarkably simple solution with $J=0,\ J' \gg \sql$ 
in the context of the AdS/CFT duality.
For example, there may exist  a more complicated
(e.g., pulsating)     solution 
with the same quantum numbers $J=0,\ J'\not=0$
whose basic features like energy dependence on $J'$ 
are  the same as of  our simple solution. 
With this motivation in mind we 
 comment on the form
of the  1-loop sigma model correction   to the energy, and
 conjecture
about    the existence of an interpolation formula for
 $E(J',\l)$.

In section 5 we discuss the structure of the  (Euclidean)  CFT
operators  which should be dual to the semiclassical two-spin
 states.
We mention, in particular, that the  1-loop  results  of
\rf{\sta,\miz} may be used to check
 our conjecture  that the anomalous  dimension of the
 scalar
SYM operator with $J=0,\ J_2=J_3 = J'$  (with lowest dimension
 above
the BPS bound)  should scale with $J' \gg 1$ as
$\D= 2J' + { \l \ov 4 J'} + ...$ not only at strong  but also at
 weak
coupling. Section 6 contains some remarks
on  generalizations and open problems.

In Appendix A we give some  details of the stability analysis
for both the $AdS_5$ and $S^5$ solutions
(this will be discussed in more detail in  \ftn).
  In Appendix B we
 derive
the quadratic fermionic part of the \adss Green-Schwarz action
 that
supplements the bosonic fluctuation actions in sections 3 and 4.
 The
total action should be  the starting point for a computation of
 1-loop
corrections  to the energies of our solutions following \fts. We
check, in particular, that the  fermionic mass matrix
 contribution to
the logarithmic divergences  cancels against the bosonic one, in
agreement with the conformal invariance of the \adss superstring
sigma model action. In Appendix C we sketch  the derivation of
 the
bosonic fluctuation actions in the conformal gauge, and check
consistency  with the static gauge results for the fluctuation
actions used in sections 3 and 4.
Appendix D  contains standard facts about relation between Young
tableau and Dynkin labels of representations of $SU(4)$
group which is used to identify the operators on the gauge theory
side.

\newsec{Two-spin solution in
  flat space and   its $AdS_5$ or $S^5$  generalizations }

\subsec{\bf Flat case}

Let us start with  closed bosonic string solutions in flat
 Minkowski
space. In orthogonal gauge, string coordinates are given by
solutions of free 2-d wave equation, i.e. by combinations  of
$e^{i n(\tau \pm  \s)}$, subject  to the standard constraints
$\dot X^2 + X'^2 = 0, \ \dot X X'=0$.  Let us consider, in
 particular,
a closed string (with its center of mass at  rest at the origin
 of
the  cartesian coordinate system)  which is rotating in the two
orthogonal spatial planes 12 and 34.    From the  closed string
equations  on a 2-cylinder  ($\tau, \s\equiv \s + 2 \pi$) with
Minkowski signature in both target space and world sheet one
 finds
\eqn\flat{
X_0= \k \tau \ , \ \ \ \ \
 X= X_1 + i X_2 = \ r_1(\s) \ e^{ i \phi(\tau)} \ ,
\ \ \ \   Y= X_3 + i X_4 = \ r_2(\s) \ e^{ i \vp(\tau)} \ ,   }
\eqn\fld{ \phi = n_1 \tau \ , \ \ \ \
\vp = n_2  \tau \ , \ \ \
\ \ \ \ r_1  = a_1   \sin (n_1 \s) \ ,  \ \ \ \ \ \
r_2  = a_2  \sin [n_2 (\s + \s_0) ]  \ .  }
Here $\s_0$ is an arbitrary integration constant, and $n_i$ are
arbitrary integer numbers. In what follows we assume that $n_i$
are positive. The relation between $a_i, n_i$ and $\k$ follows
from the conformal gauge constraint: \eqn\coop{  \k^2 = n^2_1
a_1^2 + n^2_2 a_2^2 \ . } The energy and the two spins are
\eqn\eess{ E= { 1 \ov 2 \pi \a'} \int^{2\pi}_0 d \s\ \dot X_0 = {
\k  \ov \a'} \ , }  \eqn\vop{
\ \  S_1 = { i \ov 4 \pi \a'} \int^{2\pi}_0 d \s\
( X \dot { \bar X }- \bar X \dot X)= {n_1 a^2_1 \ov 2 \a'}  \ ,
\ \ \   S_2 = S_1( X \to Y) =  {n_2 a^2_2 \ov 2 \a'}\ ,  }
i.e.
\eqn\oess{ E= \sqrt {{ 2 \ov \a'} ( n_1 S_1 + n_2 S_2 ) } \ . }
To get  the states on the leading  Regge trajectory  (having
 minimal
energy for given  values of the spins) one  is to choose
 $n_1=n_2=1$.

While for a special  solution  in \fld\ with $\s_0=0$ the string is
folded, for generic values of  $\s_0$ it has the form of an ellipse.
Another remarkable special case is when $n_2 \s_0 = {\pi\ov 2}$,
i.e.  when $|X|  \sim \sin\s$   but  $|Y| \sim \cos \s$, and
\eqn\sym{
 n_1=n_2=n\ ,\ \ \ \ \ \  \ a_1 =a_2 = {\k \ov \sqrt 2 n } \ . }
Then the string becomes {\it  circular } and,  while rotating, it
always {\it lies on $S^3$}  in $R^4$ space  formed by
$(X_1,X_2,X_3,X_4)$. Indeed, the radius  in $R^4$ then remains {\it
constant}
\eqn\spe{
 |X(\tau,\s)|^2 + |Y(\tau,\s)|^2
  = X_1^2 + X_2^2 + X_3^2  + X_4^2 = {\k^2 \ov 2 n^2 }  \ . }
 In this case  the two spins are equal
\eqn\suuo{  S_1=S_2 = {\k^2 \ov 4 n\a'} \equiv S \ , \ \ \ \ \
 E= \sqrt {{ 4n \ov \a'} S  } \ .   }
Thus, the $ X \leftrightarrow  Y$ symmetric  circular string rotating
in the  two orthogonal planes  and corresponding to a state  on the
leading Regge trajectory ($n_1=n_2 =1$)  is described by the following
solution
\eqn\syy{
X_0= \k \tau \ , \ \ \ \ \
 X_1 + i X_2 = \ {\k \ov \sqrt 2} \sin \s  \ e^{ i \tau } \ ,
\ \ \ \   X_3 + i X_4 = \ {\k \ov \sqrt 2} \cos \s  \ e^{ i \tau } \ .
  }
The crucial observation is that such solution can be  easily
generalised to a solution describing  a string rotating   in  any
homogeneous  space containing $S^3$, in particular, $AdS_n$ or $S^n$
with $n \geq  5$.

 \subsec{\bf $AdS_5$ case  }
Indeed, let us consider the $AdS_5$ space described as a hypersurface
in 6-dimensional space $R^{2,4}$:
\eqn\hypp{
X_M X_M \equiv \eta_{MN} X^M X^N =
-   X_5^2 - X_0^2 + X_1^2 + X_2^2 + X_3^2  + X_4^2   = - 1 \ . }
The corresponding string sigma model Lagrangian is ($\L$ is a Lagrange
multiplier  field)
\eqn\laga{ I = - { R^2 \ov 4\pi \a'} \int d\tau
d \s \ L \ , \ \ \  \ \ \
L = \del_a X_M \del^a X_M   + \L ( X_M X_M + 1 ) \ .}
The equations of motion in the orthogonal gauge  then are
\eqn\eqq{
  - \del^2 X_M   + \L X_M =0 \ , \ \ \ \
  X_M X_M =-1 \ , \ \ \   \ \ \  \L = \del_a X_M \del^a X_M \ , \ \ }
\eqn\coos{   \dot X_M \dot X_M + X'_M X'_M =0\ , \  \ \ \ \ \
   \dot X_M X'_M = 0 \ .  }
A special  class of solutions of these non-linear equations is
characterised by the property $\L=\const$. It is natural to organize
the six coordinates $X_M$ into the three 2-planes or complex lines,
 \eqn\coop{W\equiv  X_5 + i X_0\ , \ \ \ \ \
  X\equiv  X_1 + i X_2\ , \ \ \ \ \ \   Y\equiv  X_3 + i X_4\ , \ \ \
 \ \
 |W|^2 -  |X|^2 - |Y^2| = 1 \ . }
Then it  is easy to  check that   the following configuration is an
example of the solution of \eqq,\coos\ with $\L=\k^2 =\const$ (cf.
\syy)\foot{It would be interesting to find other solutions with
$\L=\const$. It might even be possible to classify all such
solutions.}
\eqn\taks{ W= \cosh
\r_0 \  e^{i \k \tau} \ , \ \ \ \ X= \sinh \r_0 \ \sin \s \  e^{i \w
 \tau}\
, \ \ \ \ \ Y = \sinh \r_0 \ \cos \s \  e^{i \w \tau}\ , \ \ \ \ \
}
where $\r_0$ and $\w$ are   related to $\k$ as follows
\eqn\cons{\w^2 = \k^2 + 1  \ , \ \ \ \ \ \ \ \ \ \
\sinh^2  \r_0 = \ha \k^2   \ . }
Notice that the expressions for $X$ and $Y$  look exactly the same as
in the flat space solution  \syy. Indeed, since $ |W|^2 = \cosh^2 \r_0
= 1 + \k^2 $ the $AdS_5$ constraint $ |W|^2 -  |X|^2 - |Y^2| = 1 $ is
automatically satisfied.

This solution describes a circular closed string rotating in  the 12
and 34 planes in the global $AdS_5$ time $t= \k \tau$. It will be
rederived  in the next section  using explicitly the standard set of
global $AdS_5$ coordinates $(t,\rho,  \theta,\phi,\vp)$
related to $X_M$ as follows:
\eqn\rell{
W= \cosh \r \  e^{i t } \ , \ \ \ \
X= \sinh \r \ \sin \theta \  e^{i \phi}\ , \ \ \ \ \
Y = \sinh \r \ \cos \theta \  e^{i  \vp}\  . }
It is clear that the string described by this   solution has equal
angular momenta in the 12 and 34 planes.
In this $R^{2,4}$
embedding representation it is easy to identify the charges of
the isometry group $SO(2,4)$ of $AdS_5$ that are  non-vanishing
on this solution. In general, the 15 rotation  generators
$J_{MN}$ of $SO(2,4)$  can be related to  the conformal group
generators  as follows (see, e.g., \aha) \eqn\genn{
J_{\m\n} = M_{\m\n}\ , \ \ \ \ J_{\m 4} = \ha (K_\m - P_\m)\ , \ \ \
 \  \  J_{\m 5} = \ha
(K_\m + P_\m) \ , \ \ \ \ \   J_{54} = D\  , }
where $\m,\n=0,1,2,3$.
We can identify the  standard  spin with   $S_1 = M_{12} = J_{12}$,
the second  (conformal) spin  with  $S_2 = J_{34} = \ha (K_3- P_3)$,
 and finally the rotation generator  in the $05$ plane  with the
global $ AdS_5$ energy,  $E= J_{05} = \ha (K_0 + P_0)$.\foot{After the
Euclidean continuation  $X_0\to  i X_{0E}$ and mapping to $R \times
S^3$  it is natural to classify the  representations of the conformal
group in terms of maximal compact subgroup $SO(4) \times  SO(2)$, or
$SU(2) \times SU(2) \times SO(2)$.  Exchanging  $X_{0E}$ with $X_4$
exchanges the generator  $J_{54} = D$ with $ J_{05} = \ha ( P_0 + K_0)
= E$.}  In the present case the only non-vanishing charges  are
$J_{50}$ and $J_{12}, J_{34}$,
 i.e.  the energy and the two spins \eqn\uio{ E= \sql \k
(1 + \ha \k^2)\ , \ \ \ \ \ \ \ \ \  S\equiv S_1= S_2= {1\ov 4}
\sql \k^2 \sqrt{  \k^2+1 } \ . } The  three Casimir operators of
$SO(2,4)$ are then expressed in terms of $\k$. Note also  that $
E^2 - (2S)^2 = \sql ( { 3 \ov 4} \k^4 + \k^2 )  \geq 0$. For
small $\k$ we get $E\approx  \sqrt{ 4\sql S}$, i.e. the usual
Regge trajectory relation,   while for large $\k$  we have
$E\approx 2S=S_1+S_2 $, similar to the single-spin case in \gkp.

It may be worth stressing the following point. We consider
classical string solutions with large angular momenta. In quantum
theory such a classical solution should correspond
 to a vector of an
irreducible representation labeled by the charges which would
then be (half-)integer. Since on our clasical solutions $S_1$ and
$S_2$ (and $E$) are the only nonvanishing charges among the
relevant generators $J_{MN}$, we  may  use them to label
representations of the $SO(4)=SU(2)\times SU(2)$ subgroup of the
conformal group $SO(2,4)$. Assuming that $S_1 \geq S_2$,  the
usual labels $j_1$ and $j_2$ of $SU(2)\times SU(2)$  can be
expressed in terms of  $S_1$ and $S_2$ as $2j_1 = S_1 + S_2, \  2j_2 =
S_1 - S_2$.\foot{The charges  $S_1$ and $S_2$ are,
in fact,  directly related to the Gelfand--Zeitlin labels of
representations of  $SO(4)$.}
Moreover, the fact that only $S_1$ and $S_2$ do not vanish means
that in quantum theory the corresponding quantum state should be
the highest weight vector of an $SO(4)$ representation. A similar remark will
apply to the case of solutions with angular momentum in $S^5$ we
study below: a similar relation will exist between the only
non-vanishing  $SO(6)$ charges $J_i=(J_{12},J_{34},J_{56})$ which
are directly (up to permutations) related to the Young
tableau  labels,
and the Dynkin labels of $SU(4)$ representations.

\subsec{\bf $S^5$ case }

Let us now consider an $S^5$ analogue of the flat-space solution
 \flat. Here all is similar to the  $AdS_5$ case, apart from the fact
that the decoupled time coordinate $t$ is  introduced in addition to
the $S^5$ directions $X_A$\
$$ X_A X_A = X_1^2 + ... + X^2_6 = |Z|^2 + |X|^2 + |Y|^2 = 1 \ , $$
\eqn\tikk{ Z= X_1 + i X_2\ , \ \ \ \ \ \ \ X= X_3 + i X_4 \ ,
 \ \ \ \ \ \ \ \    Y= X_5 + i X_6 \ . }
The relation to the standard 5 angles $(\g,\psi, \vp_1,\vp_2,\vp_3)$
 of $S^5$ is
\eqn\tokk{
Z= \cos  \g \ e^{ i \vp_1} \ , \ \ \ \    X= \sin   \g \ \cos \psi \
e^{ i \vp_2} \ , \ \ \ \
Y=  \sin   \g \ \sin \psi \
e^{ i \vp_3} \  . }
A particular  solution of the $S^5$ sigma model equations  (which are
the direct analogues of \eqq,\coos) is (cf. \syy,\taks)
\eqn\takk{ t= \k \tau\ , \ \ \ \ \
Z= \cos  \g_0 \ e^{ i \nu \ta} \ , \ \ \ \    X= \sin   \g_0
\ \cos \s \
e^{ i \ww \ta} \ , \ \ \ \
Y=  \sin   \g_0  \ \sin \s  \
e^{ i \ww \ta } \  ,  }
where  $\L = \nu^2$  and $\k$ and $\nu$ are independent parameters
while the constants  $\g_0$ and $\ww$ are expressed in terms of them
(cf. \cons) \eqn\hoh{
\ww^2 =1 +  \nu^2   \ , \ \ \ \ \ \ \ \ \ \
 \ \   \sin^2 \g_0 = \ha ( \k^2 -\n^2)   \ . }
Here the energy of the solution is  $E= \sql \k$,  and in addition we
have 3 non-vanishing components of the $SO(6)$ angular
momentum tensor  $J_{AB}$ :
$$ J_1 = J_{12} \ , \ \ \ \ \ \ \ \ \  J_2 = J_{34} \ , \ \ \ \ \ \ \
 J_3= J_{56}\ , \   $$     \eqn\compo{
J\equiv  J_1  = \sql \ \nu  [ 1 - \ha ( \k^2 -\n^2) ] \ , \ \ \ \
\ \ \ \ \ \ \ \
\JJ \equiv  J_2= J_3  = {1\ov 4} \sql  ( \k^2 -\n^2)
   \ \sqrt{  \nu^2+ 1  }  \ .}
This solution describes a  circular closed string  rotating (with
equal speeds)  in the two planes  in  $S^3$ within $S^5$,  with its
center of mass orbiting along the orthogonal circle of $S^5$.
When embedded into $AdS_5 \times S^5$ it will be located
at the origin $\r=0$ of $AdS_5$.  We shall return to the discussion
of this solution  in section 4.1.

Notice that \hoh\ implies the bound
\eqn\bou{  \n^2  \leq  \k^2  \leq \n^2 + 2  \ .  }
One limiting case is  $\k=\nu$  when the string is point-like and has
no spin   $J'$, i.e. moves along the  geodesic discussed in \bmn\  and
thus has $E= J$.  The other  is $\n^2 = \k^2 -2 $\  so that  $\k^2
\geq 2$. \foot{Here one cannot take  the flat-space limit in which
$\k\to 0$.}    Here    $J=0$ and the string has maximal size ($\g_0 =
{\pi\ov 2}$),  while the  energy and the two equal $SO(6)$ angular
momenta  $J_1,J_2$ take values
\eqn\kaak{
\JJ =  \ha \sql  \ \sqrt{ \k^2 - 1} \ \geq \ha\sql  , \ \ \ \ \ \ \ \
 \ \ \
 \
E= \sql \k = \sqrt{  (2\JJ)^2 + \l }\  \geq \sqrt{2 \l}   \ . }
This  expression for $E(\JJ)$ is  very simple and interesting, and we
shall return to the discussion of it in sections 4.2 and section 5.

Another special case  with $J=0$ is $\nu=0$: here $\ \k^2 \leq 2$ and
$\ww=1$  so that
\eqn\soso{
J' = {1\ov4} \sql \k^2 \ \leq \ \ha \sql\ , \ \ \ \ \ \
 E= \sqrt{ 4 \sql J'}\  \leq\  \sqrt{ 2\l}  \ . }
Remarkably, here the dependence of the energy on the $SO(6)$ spin is
exactly  the same as for the leading Regge trajectory in flat space!
This is not too  surprising since the corresponding string solution
\takk\ is then the direct embedding of the flat space solution \syy\
into  the $S^3$ part of $S^5$ (note that the Lagrange multiplier $\L$
in the $S^5$ analog of \eqq\ vanishes when $\n=0$ and so $X_M$ satisfy
the flat-space equations of motion).

\bigskip

One can also construct more general multi-spin solution  which has all
5 charges  being non-vanishing -- $S_1=S_2$ in $AdS_5$,    and $J=J_1$
and $  J_2=J_3$ in $S^5$. It will be given by  a direct combination of
\taks\ and \takk\ with the  parameters $\k,\n,\r_0,\g_0$ related by
the conformal gauge constraint as
$\k^2 = \n^2 + 2 \sinh^2 \r_0 + 2 \sin^2\g_0$.
The expressions for the $SO(2,4)\times SO(6)$ charges are  essentially
the same as in \uio\ and \compo, i.e. (see also section 4.2)
\eqn\goh{
E = \sql \cosh^2 \r_0 \ \k \ , \ \ \ \ \ \ \ \ \ \
S=  S_1 =  S_2 = \ha \sql \sinh^2 \r_0 \ \sqrt{ \k^2 + 1} \ , \ \ \ \ }
\eqn\oth{
J= J_1= \sql \cos^2\g_0 \ \nu \ , \ \ \ \ \ \ \ \ \  J' = J_2 =  J_3 =
\ha \sql \sin^2 \g_0 \ \sqrt{\n^2 + 1 } \ . }
The parameters $\r_0,\nu,\g_0$ and thus $\k$  can be expressed in
terms of $S,J$ and $J'$, giving the general expression for the energy
$E=E(S,J,J')$.

One can also study other similar multi-charge  solutions, like an
interpolation between  the $J,J'$ solution on $S^5$ and the
single-spin $S_1\not=0, S_2=0$ solution \rf{\dev,\gkp}
in $AdS_5$ (in this case the radial coordinate $\r$ will no longer  be
constant). Then one will find $E=E(S_1,J,J')$ which will be a
generalization to $J'\not=0$ of the expression obtained   in
\rf{\gkp,\fts}.

\newsec{Two-spin solution in $AdS_5$ in global coordinates
and stability
 }

Here we shall rederive the $AdS_5$
two-spin solution starting from a more
general ansatz with two unequal rotation parameters  and then study small
fluctuations near the resulting solution.

The string action
in \ads written in the conformal gauge in
terms of independent  global coordinates $x^m$ is
\eqn\gss{
I= - { \sql  \ov 4 \pi } \int d^2 \xi
 \  G^{(AdS_5)}_{mn}(x) \del_a x^m \del^a  x^n\  \ , \ \ \
\ \ \ \ \ \sql \equiv  { R^2 \ov  \aa}}
Here  $\xi^a=(\tau,\s), \ \s\equiv \s + 2 \pi$.
We shall use the  Minkowski signature  in both target space and  world
 sheet,
so that in conformal gauge $\sqrt {-g} g^{ab} = \eta^{ab}=$diag(-1,1).
The equations of motion following from the action should be
supplemented by the conformal gauge  constraints.

We shall use the following explicit parametrization of
the (unit-radius) \ads metric (related to the
embedding coordinates of the previous section
by \rell)
\eqn\add{
(ds^2)_{AdS_5}
=  G^{(AdS_5)}_{mn}(x) dx^m  dx^n
= - \cosh^2 \r \ dt^2 +  d\r^2 + \sinh^2\r \ d\Om_3 \ , }
\eqn\spp{
d\Om_3
= d \t^2 +
 \sin^2 \t  \ d \p^2 + \cos^2 \t \ d\vp^2  \ . }
 This metric  has translational isometries in $t,\phi,\vp$
 so that
 a general string  solution should
possess the following three integrals of motion:
\eqn\eee{
 E= P_t= {\sql  } \int^{2\pi}_0  {d \s\ov  2 \pi} \ \cosh^2 \r\ \pa_0 t
 \equiv \sql \E
\ ,   }
\eqn\spii{
S_1= P_\p=   {\sql  }
  \int^{2\pi}_0 {d \s\ov  2 \pi}  \ \sinh^2 \r\ \sin^2 \t \
  \pa_0 \p \equiv \sql  \ss_1  \ ,  }
\eqn\spi{
S_2= P_\vp= {\sql   } \int^{2\pi}_0 {d \s\ov  2 \pi}
 \ \sinh^2 \r\ \cos^2 \t \ \pa_0 \vp \equiv \sql \ss_2
\ . }
The first integral is the space-time energy, and the second
and third ones are the spins associated with rotations in $\phi$ and $\vp$.

\subsec{\bf Solution}

Our aim is to  look for a
 solution describing a closed
string rotating in both  $\phi$
and $\vp$, thus generalizing the single-spin solution of \rf{\dev,\gkp}.
  A natural ansatz for such a solution is
$$
t= \k \ta \ , \ \ \ \phi= \w_{\phi} \tau \ , \ \ \
 \ \ \vp=   \w_{\vp} \tau \ ,  \ \ \   \  \ \ \ \ \ \ \ \ \
 \om,\ \w_\phi,\ \w_\vp =\const \ , $$
 \eqn\sll{ \r= \r (\s) = \r(\s + 2\pi) \ , \ \  \ \
\  \  \t = \t (\s) = \t(\s + 2\pi)  \ , }
where  $\r$ and $\t$ are subject to the corresponding  second-order
 equations (prime denotes  derivative over $\s$)
 \eqn\seccr{
 \r''= \sinh\r \ \cosh \r\  (\k^2 +  \t'^2 -
\w_{\p}^2\sin^2\t -\w_{\vp}^2\cos^2\t   )\ ,  }
 \eqn\secct{
 (\sinh^2\r\ \t')' = \ha (\w_{\vp}^2 - \w_{\p}^2)\sinh^2\r \ \sin2\t \ \ .
} The   first of the   conformal gauge  constraints
\eqn\conn{
 G^{(AdS_5)}_{mn}(x) ( \dot  x^m \dot  x^n +  x'^m   x'^n)
 =0 \ , \ \ \ \ \ \ \
 G^{(AdS_5)}_{mn}(x) \dot x^m   x'^n =0 \ , }
then  implies    that $\r(\s)$ and $\t(\s)$ must satisfy  the
following 1-st order equation
\eqn\rgo{
\r'^2 + \sinh^2 \r\ \t'^2 =  \k^2 \cosh^2 \r - \sinh^2 \r\
 (    \w_\p^2 \sin^2\t   +  \w_\vp^2 \cos^2\t  )\ . }
Unfortunately, we do not know how to solve the system
of non-linear equations  \seccr,\rgo\ for generic values
of the frequencies $\w_\vp$ and $\w_\p$,
so in what follows we shall assume that
the frequencies are equal:
\eqn\same{  \w_\vp\ = \ \w_\p\ =\ \w\ .    }
Then \secct\ implies
\eqn\impl{ \theta' = { c \ov \sinh^2 \r} \ , \ \ \ \ \    c=\const.}
The special solution of \secct\  with $c=0$, i.e.
 $\theta=\const$
leads us back to the single-spin case of \rf{\dev,\gkp}:
one can make a global $SO(3)$ rotation (or redefinition of $\phi,\vp$)
 to put the rotating
string in a single plane.  If one assumes that  $\theta'\not=0$,
one can show by a detailed analysis
that there  exists no solution to \seccr,\rgo\
with non-constant $\r$, i.e.
one must set \foot{Using \impl\ we get from
 \rgo: \
$\rho'^2= - V(\r) \ , \ V(\r) \equiv
 {c^2 \ov \sinh^2\r} - \k^2 \cosh^2 \r +  \omega^2 \sinh^2 \rho$.
{}From the form of the effective ``potential'' $V$ in this equation one
finds that one cannot satisfy  the closed  string periodicity condition
in $\sigma$ \sll\  unless $\r$ is fixed to be at  zero of $V$.}
\eqn\sert{
\r(\s)=\r_0=\const\ . }
Then the equations \seccr\ and \rgo\ take the form
\eqn\seccri{
 \t'^2 = \w^2 -  \k^2\ , \ \ \ \  \ \ \ \ \ \
 \t'^2 = \coth^2\r_0\
 \k^2 -  \w^2\ .}
The solution to these equations  is given by
\eqn\soll{
\t = n\s\ , \ \ \ \k^2 = 2n^2\sinh^2\r_0\ ,\ \ \ \ \w^2 =
n^2(1+2\sinh^2\r_0)= \k^2+n^2\ .}
Here $n$ is an arbitrary integer
 representing how many times the
string ``winds'' around  the $\t$-circle (cf. \rell).
The parameter $\r_0$  determines the
radius ($\sinh\r_0$) of a circular string rotating in $S^3$.
It is remarkable that one needs two rotation parameters
to be non-zero in order to stabilize the size of the string
at fixed value of the $AdS_5$ radius $\r$.

In what follows we shall  consider the
case of (cf.\cons)
\eqn\taak{ n=1 \ , \ \ \ \ \ \ \  {\rm  i.e.} \ \ \ \  \
\k = \sqrt 2 \sinh\r_0\ ,\ \ \ \ \w^2 = \k^2+1 \ .}
In the flat space limit ($\k\to 0$,  $\r_0\to 0$)
this corresponds to a state
on the leading Regge trajectory,
i.e. having  minimal energy for a given spin.
 The dependence on the  ``winding number'' $n$
  can be easily restored in all the  equations below.

The integrals of motion \eee, \spii, \spi\  on this
 solution are given by the same expressions as in \uio\
 (we consider  the values rescaled by the  string
 tension $\sql$ as defined in
 \eee,\spii,\spi)
 \eqn\iint{ \E =\k \cosh^2\r_0
= \k ( 1 + \ha \k^2) \ , }
\eqn\integ{
\ss_1 =\ss_2 \equiv \ss\ , \ \ \ \ \ \
\ \ss =\ha \w \sinh^2\r_0
= {1\ov 4}\k^2 \sqrt{ \k^2 +1} \ , }
One can easily solve the cubic equation for $\k^2$ as a function of $\ss$
 to find the
dependence of $\E$ on $\ss$. In the case of
a small string
 with $\r_0\to 0, \k\to 0 $ (i.e. a
 string near the center of $AdS_5$) we get
\eqn\from{\E = \sqrt{4\ss}\left[ 1  +  \ss
  + O(\ss^2)\right] \ , \ \ \ \ \ \ \ \ \ \ss \ll 1 \ .  }
This  is the usual Regge trajectory relation
in flat space plus  the first correction
due to the curvature of $AdS_5$. In the case of a large
string with $\r_0\gg 1 ,\  \k\gg 1 $
(i.e. a  long  string  close  to the boundary of $AdS_5$)
we get
\eqn\iop{\E = 2\ss + {3\ov 4}(4\ss)^{{1/3}} + O({\ss^{-1/3}})\  ,
\ \ \ \ \ \ \ \ \ \ \ \ \ss \gg 1 \ . }
Note that here
the first correction to $\E-2\ss$ goes as $\ss^{{1/3}}$,
which is different from the
$\ln \ss$ behavior in the single-spin (folded rotating closed string)
 case in \gkp.
However, as  we explain in the next section,
 the solution with large $\ss$
 turns out to be unstable.

\subsec{\bf Fluctuations,  stability and 1-loop correction }

To compute the quadratic action for fluctuations near
 the above solution
it is useful to start with
the  Nambu-Goto analog of the action \gss\ and choose
 the static gauge
\eqn\jop{t = \k \tau\ , \ \ \ \ \ \ \ \ \t = \s\ .  }
Let us note  that the induced metric on our solution is flat:
\eqn\uop{
ds^2_2= \sinh^2\r_0 \  ( -d\tau^2 + d \s^2) \ . }
Shifting   the remaining three fields away from  their
classical values
\eqn\uoy{ \r\to \r_0 + \td \r\ , \ \ \ \
 \p\to \w \tau + \td \p\ ,  \ \ \ \ \   \vp\to
\w\tau + \td \vp\ , }
and expanding the Nambu-Goto action up to the second
 order in the fluctuation fields,
we get the quadratic Lagrangian for the fluctuations
 $\td \r,\ \td \p$ and $\td\vp$
$$L = - \ha (\pa_a\td \r)^2 - {1\ov 4}
\cos^2\s\ \k^2[ 1+\cos^2\s\ (1+\k^2)] (\pa_a\td\vp)^2
$$
$$
-\  {1\ov 4} \sin^2\s\ \k^2 [1+\sin^2\s(1+\k^2)](\pa_a\td\p)^2
- \ha  \cos^2\s\sin^2\s\ \k^2(1+\k^2)\pa_a\td\p\pa^a\td\vp\
$$
\eqn\lagr{
+\
2\sqrt{\k^2(1+\k^2)(2+\k^2)} \td \r(\pa_0 {\td \vp}\cos^2\s+\pa_0 {\td
 \p}\sin^2\s)
+
 (2+\k^2) \td\r^2\ .}
This Lagrangian takes  simpler form  after  making
the following change of
variables $(\td \p,\td \vp) \to (\a, \b)$ \foot{This
transformation has a very simple interpretation in terms of
fluctuations of the complex fields $X$ and $Y$ in \rell:
expanding near their classical values \taks\ in the static gauge
where $\theta$ is not fluctuating and ignoring fluctuations of
$\rho$ we get:
$\td X = i \sinh \r_0 \ \sin \s \ e^{i \w \ta} \  \td \p,
\ \ \   \td Y = i \sinh \r_0 \ \cos\s \ e^{i \w \ta} \  \td \vp$.
Then $\a,\b$ expressed in terms of $\td X$  and $\td Y$
take the form (at each given $\tau$)
of an $O(2)$  rotation with angle $\s$.}
\eqn\rety{
\a = a \ (\td\vp\cos^2\s+\td\p\sin^2\s )\ , \ \
\ \ \ \ \ \ \ \
\b = b\ \sin2\s\  (\td\p-\td\vp)\ , }
\eqn\exc{
\td\p = {\a\ov a} +{\b\ov 2 b}\tan\s\ ,\ \ \
\ \ \ \ \ \ \     \td\vp = {\a\ov a} -{\b\ov 2 b}\cot\s\ , }
\eqn\valp{
a= {1\ov\sqrt 2 \k}\sqrt{2+\k^2}\ ,\ \ \ \ \ \ \ \ \ \
\ \  b = -{ 1\ov \sqrt{2}\k}\ .  }
An  apparent singularity of the transformation
\rety\ at $\s =0,{\pi \ov
2},\pi,{3\pi \ov 2}$ is not physical because it is a reflection
of the obvious coordinate singularity of the $AdS_5$ metric 
\add,\spp\ 
at
$\theta =0,{\pi \ov 2},\pi,{3\pi \ov 2}$.\foot{While to cover 
$S^3$ once one is usually assuming $0\leq 
 \theta \leq { \pi \ov 2}$ with $\phi$ and $\vp$ changing 
 from $0$ to $2\pi$,
here to embed the closed string in $S^3$  at each fixed moment of
time  \sll\  we need to consider $\theta$ in the interval 
 from $0$ to $2\pi$.}
In terms of the new
fields \lagr\ takes the following simple form 
$$ L =- \ha
(\pa_a\td\r)^2- \ha (\pa_a \a)^2- \ha (\pa_a\b)^2 $$
\eqn\lagri{+ \ 2\sqrt{2(1+\k^2)} \pa_0 {\a}\ \td \r
-\ 2 \sqrt{2+\k^2} \pa_1 \a\ \b\
- 2(1+\k^2)\b^2+ (2+\k^2) \td\r^2 \ .}
The Lagrangian  \lagri\ can be rewritten as
(omitting total derivative)
\eqn\laak{
L = -\ha (D_a \eta^s)^2  - \ha M_{sr} \eta^s \eta^r \ , \ \ \ \ \ \ \
D_a \eta^s= \del_a \eta^s + A^{sr}_a \eta_r\ , \ \ \ \ \
\eta^s=(\td \r, \a, \b) \ , }
 $$
A_0^{\a\td \r } =-A_0^{\td\r\a}= \sqrt{2(1+ \k^2) } \ , \ \ \ \ \
A_1^{\a\b} =-A_1^{\b\a}= \sqrt{2 + \k^2 } \ , $$
\eqn\voot{M_{sr} \eta^s \eta^r
=   - 2 \td \r^2  + \k^2 \a^2  +
( 2 +  3  \k^2 ) \b^2
\ , }
where the 2-d  non-abelian  $SO(3)$ gauge field $A^{sr}_a$  has
a constant but non-vanishing  field strength $F^{\b\td\r}_{01}=
\sqrt{ 2(1+ \k^2) (2+\k^2)}$.
It is remarkable that in spite of the  $\ta$ and $\s$ dependence of our
background solution, the fluctuation action  is quite simple,
having  constant coefficients; in particular, it is
 simpler than the corresponding action in
  the one-spin case in \fts.  The absence of
explicit dependence on $\s$ will  allow us to solve the
linearized equations of motion for the fluctuations.

Note that the radial fluctuation  $\td \r$
has a negative mass term in \voot, suggesting an instability
(the Hamiltonian corresponding to \laak\ is not positive definite).
However, since it is coupled to a gauge field
(i.e. is mixing with other fluctuations) the latter
may,  in principle,  stabilize  the
$\td \r$ evolution.\foot{Examples of similar situations
are a charged  (inverted) harmonic oscillator
in a magnetic field, and a ``tachyon'' mode in $AdS$ space.}
 Thus the
stability issue needs to be carefully  studied.
This is done in Appendix A.
It is found there that
the solution is stable only for a certain range of
values of $\k$,  i.e. for  not very large
values of $\ss$ (see \integ)
\eqn\valli{
 0 \leq \k^2 \leq  {5\ov 2}\ , \ \ \ \ \ {\rm i.e.} \ \ \ \ \ \ \
 0 \leq \ss \leq {5\ov 8} \sqrt{{7\ov 2}}  \ . }
Note that for the maximal  value  of $\ss$, i.e.
 $\ss \approx 1.17$
 the value of the spin $S= \sql \ss$ is still large
 since in the semiclassical approximation it is assumed
 that  $\sql \gg 1$.

\bigskip

It is of  interest to compute  the 1-loop
 superstring sigma model correction to the
energy of the two-spin solution. In principle,
it can be done
following the same approach as was used in
\fts\ in the single-spin case.
It is straightforward to supplement
\lagri\ with the  Green-Schwarz quadratic fermionic term as in
\rf{\dgt,\fts}
(see Appendix B).  The fermionic contribution
 cancels the
2-d logarithmic divergence
 coming from the mass term in \laak\
 (which is  proportional to $\sum_r M_{rr} = 4 \k^2$).
 If we ignore the  instability of the solution for
 large $\k$  and formally  consider the limit $\k \gg 1$ in \lagri\ we will
 get
\eqn\forma{
L_{_{\k \gg 1}} \to - \ha (\pa_a\td\r)^2- \ha (\pa_a \a)^2- \ha (\pa_a\b)^2
+ 2\sqrt 2 \k \pa_0 {\a}\ \td \r
- \ 2\k \pa_1  \a\ \b\
- 2\k^2 \b^2 + \k^2  \td\r^2  \ . }
Since $\k$ is the only non-trivial parameter in \forma,
 the 1-loop correction to the energy on the 2-d cylinder
 is  expected to scale as
$\k$ (see  the discussion  \fts).
That would imply that the large $\ss$ expansion of the energy in \iop\
is corrected at the one loop order   by (cf. \integ)
$
\E_1 \sim  { \k  \ov \sql}\sim { 1 \ov \sql}
  \ss^{1/3}, \  \ss \gg 1 $.
This may be consistent with  the
following conjecture for the general behaviour of $E(S,\l)$
 \eqn\conn{
 E = 2S + [ h(\l) + f(\l)  S]^{1/3} + ...
 \ , \ \ \ \ \ \ \  S \gg 1 \ , }
 where
 \eqn\hohh{
 f(\l)_{_{\l \gg 1}}  = \l (  c_0 + { c_1\ov \sql } + ...)   \ ,
 \ \ \ \ \ \ \ \   f(\l)_{_{\l \ll 1}}  = \l (  b_0 + { b_1\l  } + ...)
   \ .      }
We caution, however, that  the instability of the solution for  large
$\k$ or large $\ss$  may preclude interpolation
from strong to weak coupling  in the large  spin  limit.
One could still hope that since the solution with large
$\ss$ should
evolve into a solution which will  still carry the same
spin, one may still
find the classical energy behaving with spin as in \iop.
However, one may not be able then to compute
the   sigma model loop
 corrections to the energy in a reliable way.

\newsec{Multi-spin string solutions in $AdS_5\times S^5$}

Let us now find  a similar rotating string  solution
in $S^5$  and its generalizations
 having spins in both $AdS_5$ and
$S^5$ factors. This was already  discussed
 in terms of the embedding coordinates
in section 2.
Here we will rederive these solutions in terms of angles of $S^5$
and study
some of their
properties in more detail.

The bosonic part of the \adss string action is
\eqn\gsss{
I= - { \sql  \ov 4\pi }
\int d^2 \xi  \ \big[ G^{(AdS_5)}_{mn}(x)
\del_a x^m \del^a  x^n\ + \    G^{(S^5)}_{pq}(y)  \del_a y^p
\del^a y^q \big] \ , \ \ \ \ \ \ \ \ \sql \equiv  { R^2 \ov
\aa} \ .} We shall use the following explicit parametrization of
the unit-radius  metric on $S^5$:
\eqn\Sd{
(ds^2)_{S^5}=   G^{(S^5)}_{pq}(y) dy^p  dy^q
= d\g^2 + \cos^2\g\ d\vp_1^2 +\sin^2\g\ (d\psi^2 +
\cos^2\psi\ d\vp_2^2+ \sin^2\psi\ d\vp_3^2)\ .}
This metric has three translational isometries in $\vp_i$,
so that
 in addition to the three $AdS_5$
integrals of motion \eee,\spii,\spi,
 a general solution should also
have  the following  three
 integrals of motion depending on the
$S^5$ part of the action:
\eqn\ji{ J_1= P_{\vp_1}= {\sql  } \int^{2\pi}_0
{d \s\ov  2 \pi} \ \cos^2 \g\ \pa_0\vp_1 \equiv \sql \J_1
\ ,   }
\eqn\jii{
J_2= P_{\vp_2} = {\sql   } \int^{2\pi}_0 {d \s\ov  2 \pi}
 \ \sin^2 \g \ \cos^2 \psi \ \pa_0 \vp_2 \equiv \sql \J_2
\ , }
\eqn\jiii{
J_3= P_{\vp_3}= {\sql   } \int^{2\pi}_0 {d \s\ov  2 \pi}
 \ \sin^2 \g \ \sin^2 \psi \ \pa_0 \vp_3 \equiv \sql \J_3
   \ .   }

\subsec{\bf Circular string rotating in $S^5$}

Let us look for a solution describing a closed string
located at the center $\r = 0$
of $AdS_5$  and at a fixed value of one of the $S^5$ angles
 $\g = \g_0 =\const$, rotating  within $S^3$ part of $S^5$
(with equal frequences as in the $AdS_5$ case),
with its center of mass orbiting
along a circle of $S^5$. A natural ansatz for such a solution is
\eqn\ann{  t= \k \ta \ , \ \ \ \  \r =0 \ , \ \ \ \g=\g_0 \ , \ \
\ \
 \vp_1= \nu \tau \ , \ \ \
 \ \ \vp_2 =  \vp_3 =\ww \tau \ ,  \ \ \ \ \  \psi= \s \ , }
 where $  \k,\g_0,  \nu,\ \ww=\const$.
The equations of motion for the fields and the
conformal gauge  constraints then
lead to the following relations between $\g_0$, $\k$, $\nu$ and $\ww$
\eqn\relS{
\ww^2 = \nu^2 +1\ ,\ \ \ \ \ \ \ \ \
\k^2 = \nu^2 + 2\sin^2\g_0\ .}
Just as  in the case of the two-spin solution in $AdS_5$ the induced
metric here  is flat
\eqn\flatt{  ds^2_2=  \sin^2\g_0\ (-d\tau^2 + d\s^2) \ . }
Taking into account that
\eqn\sert{
\J\equiv \J_1 = \cos^2\g_0\ \nu\ ,\ \ \ \ \ \ \
\J_2=\J_3=  \JJJ \ , \ \ \ \ \ \
 \JJJ\equiv   \ha\sin^2\g_0\ \ww\ ,  }
we find the following equation for $\nu=\nu (\J,\J')$
\eqn\nuj{\nu \sqrt{\nu^2+1}=
\J\sqrt{\nu^2 +1} + 2\JJJ \nu   \ .}
Since the energy $\E$ in \eee\ is equal to $\k$,
we can use eq.\nuj\ to find the dependence of the energy on the
R-charges $\J$ and $\JJJ$
\eqn\gty{
\E^2 = \nu^2 + {4\JJJ\ov\sqrt{\nu^2+1}} \ , \ \ \ \ \ \ \ \ \ \ \
\E=\E(\J,\JJJ) \ .  }
It is instructive to restore the $\l$-dependence in the formulas
\nuj\ and \gty, i.e. to rewrite them in terms of the energy $E$,
 the R-charges $J,\ J'$ and the auxiliary ``charge''
 ${\II} = \sqrt{\l}\ \nu= \cos^{-2} \g \ J$:
\eqn\nuJ{
\II \sqrt{\II^2+\l}=
J\sqrt{\II^2 +\l} + 2J'\II  \ ,\ \ \
\ \ \ \ \ \ \    \II\equiv \sqrt{\l}\ \nu\ ,}
\eqn\gtE{
E^2 = \II^2 + {4\l J'\ov\sqrt{\II^2+\l}} = \II^2  +
2\l (1-{ J\ov \II })  \ , \ \ \ \ \ \
\ \ \ \ \ E=E(J,J') \ .  }
A nice feature of this representation
 is that if $\II \gg\sql$ then the
expression for the energy takes the form of a perturbative
expansion in $\l$ because $\II$ can be found from \nuJ\
 as a series
in $\l\ov {(J+2J')}$. In particular, we get the
following expression
for the energy at the first order in $\l$
\eqn\rty{
E^2 \approx (J + 2J')^2 + {2\l J'\ov J+ 2J'}\ , \ \ \ \ \ \ \ \
E \approx J + 2J' + {\l J'\ov (J+ 2J')^2}\ , }
where $\II \approx J + 2J' - {\l J'\ov (J+ 2J')^2}$.
Note that here  there is no restriction on values of $J$ and $J'$
 apart from the requirement
 that $J + 2J'\gg\sql$.

 One may be
tempted  to conjecture that the formula  \rty\
may be  valid at {\it small } values
of $\l$ if $J + 2J'$ is very large. However,
\rty\ was obtained in the  strong coupling $\l\gg 1$
 regime, and we
expect  it to  receive $1\ov \sqrt{\l}$ string sigma
model corrections. In particular, even the coefficient in front
of $J+2J'$ may get changed by the corrections, and, if so,
the one-loop perturbative correction to the dimension of the
corresponding CFT
operator dual to the string solution will not  be  of order
$J'\ov {(J+2J')^2}$ but of  order $J+2J'$.

If $J\gg J'$ the energy \rty\ takes the form
\eqn\rty{
E \approx J + 2J' + \l{J'\ov J^2}\ . }
This expression for $E$ is consistent with  the  string oscillation
spectrum  in the sector with large
$J\gg\sql $ \rf{\bmn,\mets},
i.e. with the plane-wave spectrum (similar
 comparison was done in
\fts).  From the plane-wave spectrum point of view,   $J'$
represents the angular  momentum carried by string  oscillations.
Since the linear term in $J$ is not renormalized in the BMN
limit, one may conjecture that the same  should happen here.

If we set $J=0$ in \rty\  we get
\eqn\tyi{
E^2 = (2J')^2 + \l\ , \ \ \ \ \ {\rm i.e.}\ \ \
\ \ \ \ \ E \approx 2J' + {\l\ov 4J'} \ .  }
Thus at large $J'$ the correction goes as $1\ov {J'}$,
instead of a
constant shift found in the case of the single-spin folded
string rotating in $S^5$ \gkp.

We can also consider the case with $\II \ll \sql $
\ (when $\II\approx J$)
\eqn\oyp{E^2 \approx 4\sqrt{\l}J' +J^2 \ .  }
Setting $J =0$ we reproduce the usual
 Regge trajectory  relation.

It is interesting to  note  that the limit $J\to 0$ depends on the value
of the second angular momentum  $J'$ (see also the discussion
of this case in sections 2 and 4.3). When $J' =\ha\sqrt{\l}$ the
dependence of the energy on the  angular momentum  changes its
form, i.e. the system undergoes a kind of ``second order phase
transition''  (the second derivative of the energy over the
orbital momentum  has a  discontinuity  at $J' =\ha\sqrt{\l}$). This
happens because  for $\nu=0$    one has $\JJJ= \ha\sin^2\g_0$, so
that the value  $\JJJ=\ha$  is found when the  string reaches its
maximal  size ($\g_0 = {\pi\ov 2})$, i.e. when
 it  rotates on the maximal-size 3-sphere  within $S^5$.

\subsec{\bf  Circular string rotating  in both
 $AdS_5$ and $S^5$}
\bigskip

As already discussed in section 2, it is straightforward
to  combine the two-spin solution in $AdS_5$
with the three  angular  momenta solution in $S^5$.
For completeness, let us  summarize the resulting solution
depending on 3 different parameters in terms of the
global coordinates of $AdS_5$ and $S^5$ used in section 3 and
in this section:
\eqn\hop{ \r = \r_0\ ,  \ \ \ \ \ \  t= \k \ta \ ,\  \  \
\ \ \ \vp=\p =\w \tau \ ,\ \ \ \ \  \t=\s \ , }
\eqn\tyo{
\ \  \g = \g_0\ ,\ \  \   \
\vp_1= \nu \tau \ , \ \ \vp_2 =  \vp_3 =\ww \tau \  ,  \ \ \ \ \ \
\psi= \s\ . }
The equations of motion and the conformal gauge
 constraint lead to the following
relations
\eqn\req{
\w^2=\k^2 + 1\ ,\ \ \ \ \
 \ \ww^2=\nu^2 + 1\ ,\ \ \ \ \ \ \ \  \ \k^2 =\n^2 +  2\sinh^2\r_0 + 2
\sin^2\g_0 \ ,   }
and
the energy and the 5 conserved charges are given by
the same relations as in \goh,\oth\
\eqn\fop{
\E = \k\ \cosh^2\r_0 =\ \k\ [1 + \ha (\k^2 -2\sin^2\g_0 - \nu^2)]\ ,}
\eqn\fou{ \ss =\w \sinh^2\r_0 = \ha(\k^2 -2\sin^2\g_0 - \nu^2)\sqrt{\k^2 +
 1}
\ ,}
\eqn\gou{ \J = \cos^2\g_0\ \nu\ ,\ \ \ \ \ \ \ \
\ \JJJ = \ha\sin^2\g_0\ \ww=
\ha\sin^2\g_0\sqrt{\nu^2 + 1}\ .  }
One can use these equations to analyse the dependence of the
energy on the spins and $SO(6)$ charges. In particular,  when
$\J$ is very large while the string size is small,  one
reproduces the corresponding part of the oscillator plane-wave
string spectrum (with spin $\ss$ and angular  momentum $\JJJ$
here carried   by the semiclassical string instead of    being
produced by string   oscillations as in   \bmn).

\subsec{\bf Fluctuations and stability of $S^5$ solution
 with $J=0, \ J'\not=0$}

Let us now  discuss the stability of the simplest
$S^5$  solution  with $\J =0$ and $\JJJ\not=0$.
There
 are two different cases that should be discussed separately.
\bigskip

\noindent {\bf  $\JJJ \leq  \ha$  case}

The solution with $\JJJ \leq \ha$ is found
by setting $\nu = 0$ in \relS,\sert\ (see also \soso).
Then  (see \oyp)
\eqn\thenn{
\n=0\ , \ \ \ \ \
\ww = 1\ ,\ \ \ \ \  \k^2 = 2\sin^2\g_0  \leq 2 \ , \ \ \ \
\JJJ= {1\ov 4} \k^2 \ , \ \ \ \ \  \E =\sqrt{4\JJJ} \ . }
As was already  mentioned in section 2  (below \soso),
 this
solution is essentially the embedding of the flat
space solution into
$S^5$. However, the fluctuation  spectrum will of course be different
from the flat space case.

The computation of the quadratic fluctuation
 action is a  repetition of
  the one done in the  $AdS_5$ case in
section 3.2. We choose the static gauge
$$t = \k \tau\ ,\ \ \ \ \ \ \ \ \psi=\s$$
and
expand the Nambu-Goto action up to the second order in fluctuations.
As in the case  of point-like string orbiting
in  $S^5$ \fts,
the  Lagrangian for the $AdS_5$ fluctuations will be represented by
the 4  massive field  contributions\foot{Expanding near the
 $\r=0$ point in \add\ one needs to introduce the 4 cartesian-type
coordinates, e.g., writing the $AdS_5$ metric as
$ds^2 =
-  { (1+\fo \eta^2 )^2 \ov (1-\fo \eta^2 )^2  }  dt^2 +
  {  d\eta_k d\eta_k   \ov  (1- \fo  \eta^2   )^2 } $.}
\eqn\adsd{
L_{AdS_5} = - \ha ( \del_a\td  \eta_k)^2
- \ha \k^2 \td \eta_k^2  \ , \ \ \ \ \ \ \ \ \ \ \ \
k=1,2,3,4 \ .  }
The additional contribution of $S^5$ fluctuations is (cf.\lagri)
$$
L_{S^5} =- {1\ov 2}(\pa_a\hat \vp_1)^2 -
\ha (\pa_a\td \g)^2- \ha (\pa_a \a)^2- \ha (\pa_a\b)^2-
2\m \pa_0{\a}\ \td \g
$$
\eqn\lagrS{
-  2 \sqrt{2} \pa_1 \a\ \b + \ \m^2  \td\g^2 - 2\b^2\ , \ \ \ \  \ \ \
\ \ \ \   \m^2 \equiv 2-\k^2 \ .}
Here  $\hat \vp_1 =  {1 \ov \sqrt 2} \m \td \vp_1$
and the fields $\a$ and $\b$ are defined as in \rety\
\eqn\fluu{
\a = -\k (\td \vp_2\cos^2\s+\td \vp_3\sin^2\s )\ , \ \
\ \ \ \ \
\b =-{\k \ov 2\sqrt{2}}\sin2\s \ (\td\vp_2-\td\vp_3)\ .   }
This  Lagrangian is  similar to the one \lagri\ for
fluctuations around the two-spin solution in $AdS_5$.
  Its $\td\g, \a,\b $ part
can be written  in the form  \laak\ as follows
$$
L(\td \g,\a,\b) =
  \ha (\pa_0 \td \g + \m \a)^2-  \ha (\pa_1 \td \g )^2
 +   \ha (\pa_0 \a - \m \td \g)^2-  \ha (\pa_1 \a  + \sqrt 2  \b )^2
 $$ \eqn\writ{
  +\    \ha (\pa_0 \b)^2-  \ha (\pa_1 \b  - \sqrt 2  \a )^2
 + \ \ha \m^2 \td \g^2  + \ha (2-\m^2)  \a^2
  - \b^2 \ .  }
  Note that the sum of squares of masses  here vanishes,
  in agreement with the discussion of fluctuation Lagrangian
  in conformal
  gauge in Appendix C.

The Hamiltonian corresponding to \writ\  does not appear to be
positive definite, and so the stability  of the solution is  a
priori in question. It is shown in Appendix A that the solution
is, in fact,  stable  if
$${1\ov 2}\le \mu^2 \leq 2\ ,\ \ \ \ \ {\rm i.e.}\ \ \ \ \ \ \ \ 
\ \ 0 \le \J'
\leq {3\ov 8}\ .$$

\bigskip

\noindent {\bf  $\JJJ \geq  \ha$ case}

The solution with $\JJJ \geq  \ha$ is found by setting $\g_0 = {\pi\ov 2}$.
Since $\cos \g_0 =0$, the value of $\nu$ in this case  is undetermined
(cf. \Sd),  and the conformal gauge constraint gives (cf. \relS,\sert,\tyi)
\eqn\taq{
\k^2 = \ww^2+1  \geq 2 \ , \ \ \ \
  \JJJ = \ha\ww \ , \ \ \ \  \E =\k \ , \ \ \ \ {\rm i.e. } \ \ \ \
\    \E  = \sqrt{ (2\JJJ)^2+1 }\ . }
 The coordinates $\g$ and $\vp_1$ are not suitable  for studying
fluctuations around $\g ={\pi\ov 2}$ (which is a center of the
``2-sphere'' part $d\g^2 + \cos^2\g\  d\vp_1^2$  of
the
 $S^5$ metric  \Sd).
Introducing instead  the ``cartesian-type'' coordinates $X_1,X_2 $ as in
\tokk\ (which have zero values on the classical solution)
 \eqn\jol{
 Z=  X_1+iX_2 = \cos\g\  \ee^{i\vp_1}\ , }
we find that the quadratic fluctuation  action (obtained in the static
gauge  $t= \k \ta, \ \psi=\s$)   is then given by the sum of the $AdS_5$
part \adsd\  and (we ignore total derivative terms)
\eqn\sph{
L_{S^5}  = -
{1\ov 2}|\pa_a Z|^2- \ha (\k^2-2)|Z|^2 + L(\a,\b) \ , }
$$ L(\a,\b)=
- \ha (\pa_a \a)^2- \ha (\pa_a\b)^2  - 2\k \pa_1 \a\ \b -
 2(\k^2-1)\b^2
$$ \eqn\aba{
= \  \ha (\pa_0 \a)^2 - \ha (\pa_1\a  + \k \b)^2
+  \ha (\pa_0 \b)^2 - \ha (\pa_1\b  - \k \a)^2
+ \ha \k^2 \a^2 - \ha ( 3 \k^2 -4) \b^2  \ , }
where,  as in  \rety,\fluu,
 \eqn\fiii{
\a = -\k (\cos^2\s\ \td \vp_2 +\sin^2\s\ \td \vp_3 )\ , \ \
\ \ \ \ \ \ \ \
\b =-{1\ov 2}\sin2\s\ (\td\vp_2-\td\vp_3)\ .}
The  Lagrangian \aba\ is simpler than \lagrS, describing a collection of
2 coupled fields with a constant   abelian connection;
 however, the mass
matrix   is not $O(2)$ invariant, so  after the rotation there will be a
remaining $\s$-dependence in the mass matrix.   Note that the sum of
mass-squared terms  for $S^5$ fluctuations is equal to $2 (\k^2 -2)  -
\k^2  + 3 \k^2 -4  = 4(\k^2 -2)$  in correspondence  with the results in the
conformal gauge  and  with the cancellation of divergences between the
bosonic and  fermionic sectors (see Appendices B and C).

The negative mass term for $\a$ in \aba\
 raises again the question about
stability.
 To analyze the stability of this
 solution it is sufficient to
consider only the $\a,\b$ part \aba\ of the Lagrangian.
Following the
procedure explained for the $AdS_5$ case in Appendix A,
i.e. expanding the
fluctuations $\a$ and $\b$ in Fourier series in $\s$, and then  looking for
solutions in the form $\ee^{i\w_n\tau}$, we
find the following frequency spectrum
\eqn\specSi{
\w_n^2 = n^2 + 2(\k^2 -1) \pm  2 \sqrt{(\k^2-1)^2 +\k^2n^2}\ . }
It is clear that the $\omega_n$ 
spectrum is real if
\eqn\ppu{
[n^2 + 2(\k^2 -1 )]^2 - 4[(\k^2-1)^2 +\k^2n^2]
= n^2(n^2-4) \geq 0\ .}
This condition does not depend on $\k$ and is not 
satisfied for the mode with 
$n=\pm 1$.\foot{The stability is obvious for the $e^{in\s}$ 
fluctuation modes with  $n\geq 2$.
Indeed,
 \aba\ can be written as
$L(\a,\b)= \ha (\pa_0 \a)^2 - \ha (\pa_1\a  + 2 \k \b)^2
+  \ha (\pa_0 \b)^2 - \ha (\pa_1\b )^2  + 2 \b^2 $,
and thus the corresponding Hamiltonian is non-negative
for the modes with  $n\geq 2$.
Let us note  also that the  problem of analysing the
small fluctuation spectrum
of this theory is similar to the one of solving
string theory  in ``non-diagonal'' (metric and 2-form field)
 plane-wave background (cf. \papd). } 
 A possible interpretation 
  of this mode is that  
 in the frame rotating
 together with the string where string is at rest, 
 it describes  the obvious 
 instability of a circular string wound around large circle
  of $S^3$  \ftn.\foot{Such unstable mode 
  would be absent in the  $S^5/Z_2$  case.}
 
 We conclude that the rotating solution with $J=0$ 
is not stable  for any value of the angular momentum $J'\geq
\ha \sql $.

\bigskip

As was already mentioned in the introduction,
to get a stable solution with large $J'$ one needs also 
to switch on a non-zero (and large) value of the angular momentum 
$J$  \ftn.
If one could ignore the  instability, 
the solution with $J=0, \ J' \geq \ha \sql$
 would be  the  most simple and interesting case
 for the study of the AdS/CFT correspondence in a novel sector of
 states.
  One could try to compute 
 the 1-loop string sigma model  correction
 to the classical energy in \taq\ by starting with the sum of the
 bosonic fluctuation action \aba\ 
 (assuming one could formally 
 omit the unstable mode absent at 
 large $J$) 
  and the fermionic action
 derived in Appendix B.
 This will be discussed  (for the general stable 
 case of $J,J'\not=0$) in 
 \ftn.
  To estimate this correction at large values of  $\JJJ$
 one may note
 that for  large $\k$  all non-zero masses
 of the 2-d fields  are equal to  $\k$,
 and thus (see \fts) the
1-loop correction to the energy  should be expected to
scale as
\eqn\onm{
\E_1 \sim {\k\ov \sql} \sim {1\ov \sql} \JJJ \ , \
 \ \ \ \ \   \ \ \  \JJJ\approx \ha \k \gg 1 \ .
 }
 That seems to  suggest
the following interpolation formula for the energy (cf. \tyi)
\eqn\pyo{
E= 2 h(\l) J' + f(\l) { \l\ov 4J'}  \ , \  \ \ \
\ \ \ J' \gg  1  \ , }
\eqn\truo{
h(\l)_{_{\sql \gg 1} } = 1 + {a_1 \ov \sql} + ...
\ , \ \ \ \ \ \ \ \  f(\l)_{_{\sql \gg 1} }
= 1  + {b_1 \ov \sql} + ...  \ . }
In spite of the absence of 2-d supersymmetry
in the corresponding quadratic part of GS action, it may
actually happen that  the coefficients $a_1$ and $b_1$
vanish, i.e.  the first two terms in the energy
are not
renormalized at the leading order
in $1\ov \sql$ expansion.
That would  support the conjecture,
prompted  by the appearance of the first power of $\l$
in the calssical expression for $E$, that
$E= 2 J' +  { \l\ov 4J'} + ...$  is actually
true also at weak coupling, i.e. is the exact result
for the first two terms in the
anomalous dimension of the corresponding dual
 gauge theory operator.
 We shall discuss this conjecture further in the  next section.

\newsec{Towards testing
AdS/CFT duality in non-supersymmetric
  multi-spin
sectors}

Let us now discuss the gauge-theory operators that should
 correspond
to the string states represented by the classical solutions
found above. The eventual
aim  is  to try to compare their anomalous
dimensions as functions of spins and R-charges to the
semiclassical expressions for the energies found above.

\subsec{\bf $AdS_5$ rotation  case}

The semiclassical closed  string states found in global coordinates
in
$AdS_5\times S^5$  should be dual to  SYM states  on $R \times
S^3$.
Going through the usual argument of Euclidean continuation and
conformal mapping to $R^4$ (cf. \oog)  they should correspond
to local  operators in Euclidean 4-d space. One can then rotate
back to
Minkowski space, but here we prefer not to do that.
The $SO(4)$ isometry of $S^3$ in $AdS_5$ is then becoming the
``Lorentz'' symmetry of $R^4$. Thus the Euclidean
 gauge theory operators  will be classified  by its
 representations,
 i.e. will be labelled by the values $(S_1,S_2)$
 of the two $SO(4)=SU(2) \times SU(2)$ spins. In addition, they
 will
 carry also the three  quantum numbers of $SO(6)$  R-symmetry
 group.

 Let us first recall the form of the gauge theory operators that
 are
 expected  to be dual to the single-spin string state in $AdS_5$
\gkp. If $\P_M$\  ($M =1,...,6$) are the adjoint scalars  of $N=4$
SYM
theory and $D_\m$ is the covariant derivative,  a representative
operator
with canonical dimension $\D_0= S+2$ is  the standard
gauge-invariant
minimal twist operator
$ O_{\{\m_1 ... \mu_S\}}  =  \tr\left(\P_M D_{\{\m_1 } ...
D_{\m_S\}}
\P_M\right)$, where ${\{\m_1 } ...{\m_S\}}$ denotes symmetrization
and subtraction of traces. This operator will in general mix with
similar operators obtained by replacing the scalars $\P_M$ by the
gauge field strength  $F_{\m\n}$ or by the fermions so to find
its perturbative anomalous dimension one would need to
diagonalise the corresponding anomalous dimension
matrix (see, e.g., \rf{\dol,\lip,\gor}  and references
 there).\foot{Two-loop anomalous dimensions for some higher
spin currents were found in \arut.}
  The equivalent form of the
above operator is  \eqn\qqp{  O_S= \tr\left(\P_M D_{\m_1 } ...
D_{\m_S} \P_M\right)\ n^{\m_1} ... n^{\m_S}  \ =\  \tr\left[
\P_M (n^\m  D_\m)^S \P_M\right]  \ , }
 where the multiplication by the product of constant null vector
 $n^\m$
 ($n^{\m} n_{\m} =0$)  factors
 implements the symmetrization
and subtraction of traces. In Minkowski $R^{1,3}$ theory one may
choose
$n^\m= (1,0,0,1)$,  getting
$O_S=  \tr\left(\P_M (D_+)^S  \P_M\right) $, \ $D_+ = D_0 + D_3$.
In the Euclidean version which we  use here
one is to choose $n^\m$ to be complex,  e.g.,
$n^\m= (1,i,0,0)$, getting
\eqn\sto{
O_S=  \tr\left[\P_M (D_X)^S  \P_M\right] \ , \ \ \ \ \ \ \ \
\ \ D_X \equiv  D_1
+ i  D_2 \ . }
It is now clear how to generalize this discussion to the case of
operators carrying two spins of $SO(4)$: a representative operator
will be
\eqn\stoo{
O_{S_1,S_2}=  \tr\left[\P_M  (D_X)^{S_1}  (D_Y)^{S_2} \P_M\right] \
,
 }
 \eqn\gag{  \ \ D_X \equiv  D_1 + i  D_2\ , \ \ \ \ \
 D_Y \equiv  D_3 + i  D_4\ , \ \ \ \ \ \ \
 \Delta_0 = S_1 + S_2 + 2
\ . }
Since the covariant derivatives $D_X$ and $D_Y$  do not commute,
this operator  will be mixing, in particular,
with  various other operators containing
permutations of $S_1$
derivatives $D_X$ and $S_2$ derivatives $D_Y$
and having the same canonical
dimension, e.g.,
\eqn\too{\tr\left[\P_M  (D_X )^{k_1} (D_Y)^{m_2}  ... (D_X)^{k_l}
 (D_Y)^{m_l}
\P_M\right]\ , \ \ \  \ \ \ \
 \sum_i k_i = S_1\ , \ \ \ \sum_i m_i = S_2 \ . }
The eigenvector of anomalous dimension matrix
is expected to be a particular
combination of such operators (in addition to others involving
gauge field strength and fermions).
An irreducible representation of the rotation group is
represented by a particular Young tableau  with two rows with
$S_1$ and $S_2$ as numbers of boxes, i.e. should contain additional
antisymmetrizations of $D_X$ and $D_Y$ factors in \stoo.\foot{Due
to these
 antisymmetrizations the resulting  operators may not be
 superconformal
primary operators.}

The equivalent ``covariant'' form of \stoo\ is  found by
introducing two  independent null vectors $n^\m$  and $m^\n$ and
generalising \qqp\ as follows
\eqn\too{
O_{S_1,S_2}=  \tr (\P_M D_{\m_1 } ... D_{\m_{S_1}}
D_{\n_1 } ... D_{\n_{S_2}}
\P_M) \   n^{\m_1} ... n^{\m_{S_1}}  m^{\n_1} ... m^{\n_{S_2}}
  \ , \ \ \ \
 n^\m n_\m = 0 \ , \ \   m^\m m_\m =0 \ .  }
This operator   can be readily
extended to the Minkowski version  of the theory  by choosing,
e.g.,
$n^\m= (1,0,0,1)$  and $m^\m = ( 0,1,i,0)$  with
 $\eta_{\m\n} = (-1,1,1,1)$.

As is well known \lip, for  $S_1=S\gg 1 , \ S_2=0$ (or vice
versa) the  perturbative anomalous dimension of such operators
scales as $\ln S$, which is the same as the scaling of
 the energy of the single-spin rotating string
solution in $AdS_5$;  this  strongly supports the existence
of  an interpolation formula $\D=S + f(\l) \ln S$ between
the weak coupling and the strong
coupling regions  \rf{\gkp,\fts}.

To compare with the string solution found in the present paper
where $S_1=S_2= S  \gg 1$  one needs to know the perturbative
(one-loop) anomalous dimension of the operators like
\eqn\stoo{
O_{S,S}=  \tr\left[\P_M (D_X)^{S}  (D_Y)^{S} \P_M\right]  + ...
\ . }
where dots stand for appropriate permutations of $D_X$ and $D_Y$.
We are not aware of computations
  of anomalous dimension
of such operators  in the literature, and here can only speculate
 about possible interpolation formula for $\D(S)$ in this case
 (see also section 3.2). Assuming one can trust the expression
 \iop\ for the energy for large $\ss= {S\ov \sql}$
 (in spite of instability of the solution for large $\ss$)
 one would expect to find the order $S^{1/3}$ correction in the
 anomalous
 dimension replacing the familiar
 $\ln S$ correction for the single-spin operators
 \sto.  However, it seems hard to imagine how such fractional-power
 term
 could appear in the 1-loop SYM computation for the anomalous
 dimension
 of \stoo. We suspect that the interpolation formula \conn,\hohh\
 may be
 a more plausible alternative, implying that at large $S$ but small
 $\l$
(with $ \l S \ll 1$)  one should expect to find the anomalous
dimension
of \stoo\  going as
\eqn\exx{
\D =  2 k(\l) S + ... \ , \ \ \ \ \ \
 k(\l) = 1 + a_1 \l + a_2 \l^2 + ... \ . }
 It would be very interesting to
 check this by direct perturbative computations  on
 the  gauge theory  side.

\subsec{\bf $S^5$ rotation  case}

The construction of operators that carry several $SO(6)$
``spins'' and thus  should be dual to the string states
represented by the solutions in section 4 describing
  closed strings rotating in $S^5$  is somewhat similar.
Let us introduce the notation  for the three complex scalars of
$N=4$ SYM theory (cf. \tikk)
\eqn\nott{
\P_Z= \P_1 + i \P_2 \ , \ \ \ \ \ \  \P_X= \P_3 + i \P_4 \ ,
\ \ \ \ \ \   \P_Y= \P_5+ i \P_6 \ . }
Among the operators with the minimal canonical dimension for given
$(J_1, J_2,J_3)$ values of $SO(6)$ charges there are operators with
holomorphic dependence of the three complex scalars:
\eqn\trw{
O_{J_1, J_2,J_3} =
 \tr [ (\P_Z)^{J_1} (\P_X)^{J_2}(\P_Y)^{J_3}] + ... \ ,
 \ \ \ \ \    \D_0 = J_1 + J_2 + J_3 \ , }
where dots stand for permutations  of $\P_X,\P_Y,\P_Z$
factors needed to form an irreducible representation of $SO(6)$
that is expected to  be an eigenvector of the anomalous dimension
matrix. The 1-loop anomalous dimension matrix for generic scalar
operators of the form
\eqn\genn{
{\cal O}_{M_1....M_j} = \tr ( \P_{M_1} .... \P_{M_j} )  }
was computed in \rf{\sta,\miz}.
Taking its symmetric traceless part
(i.e. multiplying \genn\ by a null vector, e.g.,
$n^M=(1,i,0,0,0,0)$) one finds a chiral primary operator whose
dimension is protected. This case is equivalent to \trw\ with
$J_1=J$ and $ J_2=J_3=0$, i.e. $O_{J,0,0} = \tr  (\P_Z)^{J}$,
which is dual to the  ground state  of the string  theory
expanded near the point-like  string orbiting in $S^5$
\bmn.\foot{We are considering only single-trace operators as
seems appropriate for the elementary string state -- gauge
theory operator correspondence in the large $N$ limit. Note that there exist also
multi-trace operators  which may carry the same quantum numbers
and may be 1/4 or 1/8 BPS (see, e.g., \hok\ and refs. there).}

The operator that should be dual to the string solution with
$J_1=J, \ J_2=J_3= J'$ found  above  should then be
\eqn\ert{
O_{J, J',J'} =
 \tr [ (\P_Z)^{J} (\P_X)^{J'}(\P_Y)^{ J'}] + ... \ ,
 \ \ \ \ \   \ \ \ \   \D_0 = J + 2J' \ .  }
This  operator belongs to the irreducible
representation of $SU(4)$ with Young tableau  labels
$(J,J',J')$ or with Dynkin labels  $
[0,J-J',2J']$ (see Appendix D)\foot{We are grateful to
M. Staudacher and N. Beisert  for  correcting a mistake
in this identification in the original version of this paper.}
 if $J\geq J'$,
 and to the representation
$(J',J',J)=[J'-J,0,J'+J]$  if $J'\geq J$,
and does not seem to be a superconformal primary operator. For
example, it is known that the operator with $J=0,\ J'=2$ is a
superconformal descendant of the Konishi operator $K = \tr\left(
\P_M \P_M\right)$. In the near-BPS limit  $ J \gg J'$ the
operators of the form \ert\ are examples of the BMN  operators
\bmn\ with a small number of impurities, and one can,  in
principle, make detailed comparison between semiclassical
predictions \rty\ and the perturbative results  of \bmn\ and
\rf{\sta,\miz}. If $ J'$ is comparable to or   much larger than $J$, we
are very far from the BPS operator $ \tr (\P_Z)^{J}$,  and the
conformal dimensions of the operators cannot be computed  from
the plane-wave string spectrum.

The semiclassical results
obtained in section 4 are the only source of nonperturbative
predictions for the dimensions of these operators.
 Let us stress again  that among a
large number of operators in  these representations only
the one with the lowest conformal dimension
should be  dual to the string
solution we found.
It is also  interesting to point out
 that one and the same
formula \rty\ should be giving  the
 conformal dimensions of the operators from the
two different ($[0,J-J',2J']$ or
 $[J'-J,0,J'+J]$) representations. This
  should be true  not only in
the large $\l$ limit but also  in
the weak-coupling  perturbation theory.

As discussed in section 4.3, the simplest novel case for a
non-trivial check of the  AdS/CFT duality in a non-supersymmetric
sector is when  the circular  string orbiting in $S^5$ has
maximal size, i.e.  has $J=0$,  $J'\geq \ha \sql $. The exact
expression for   its classical energy given in \kaak,\tyi\  is
$E = \sqrt{ (2J')^2 + \l} $. 
Provided the instability of this solution could  be  
  ignored   (e.g., by embedding it into a class 
of stable solutions with $J\not=0$) 
for 
 large $\JJJ = { J'\ov \sql}$
the expression for the energy $E(J')$ 
 gives the following prediction for the strong-coupling
behaviour  of the anomalous dimension  of the corresponding
operator
\eqn\opr{ O_{0,J',J'} =
\tr [ (\P_X)^{J'}(\P_Y)^{J'}] +... \ ,  }
\eqn\predd{
\D= 2J' + { \l \ov 4 J'} + ... \ , \  \ \ \ \ \ \   \  J' \gg \sql
\gg 1 \ .
\ }
As was already discussed  in section 4.3,
our conjecture is that this expression is actually valid also at
{\it  weak }   coupling, i.e. the first two terms in $\D$ are
 not renormalized.

One can  check this  against the results of \bie\ applied to the
operators transforming in the $[J',0,J']$ representation for
$J'=4$ and $J'=5$. At $J'=4$ one finds that the
lowest anomalous dimension of the operators in $[4,0,4]$ is
$\g = 0.0411\l$ while eq.\predd\ predicts the anomalous dimension
to be $\g = 0.0625\l$, i.e. the deviation is about 34\%.\foot{
Diagonalizing the matrix of anomalous dimensions from
\rf{\sta,\miz}
one finds also an operator  with much closer value
$\g \approx 0.0619\l$, but  comparison  with the results of
 \sta\
shows that it belongs to [2,4,2] representation.
We are grateful to G. Arutyunov and J. Minahan  for explanations
related to this point.} However, at $J'=5$ one gets the lowest
anomalous dimension to be $g = 0.042\l$ while eq.\predd\ predicts
$\g = 0.05\l$, i.e. the deviation in this case is just 16\%.
It would be useful to compute one-loop anomalous dimensions of
the $[J',0,J']$ operators for $J'=6$ to see if the agreement with
eq.\predd\ is really getting better at large $J'$.

Since it is sufficient to consider the  operators with  holomorphic
dependence on the  fields $\P_X$ and $\P_Y$,  the Hamiltonian of
the integrable $SO(6)$ spin chain considered in \miz\ reduces to
the Hamiltonian of the simplest XXX$_{1/2}$ spin model \fadd.
It would be very interesting to find the corresponding one-loop
anomalous dimension and the explicit form of the  associated
operator \opr\ in the large $J'$ limit by utilizing this
connection to the XXX$_{1/2}$ model.

\newsec{Concluding remarks  }

There are various  generalizations of  the $AdS_5$ and $S^5$
solutions we have found.  For example, the  special $S^5$
solution considered in section 4.3 with  maximal size of the
string ($\g_0={\pi\ov2}$) is also a 
solution of string theory
on $R_t \times S^3$:
$$
ds^2 = - dt^2 + d\psi^2 +
\cos^2\psi\ d\vp_2^2+ \sin^2\psi\ d\vp_3^2\ ,
$$ \eqn\truu{
t=\k \ta \ , \ \ \ \psi=\s \ , \ \ \ \
\vp_2 =\vp_3 =\ww\ta \ , \ \ \  \k^2 = \ww^2 + 1 \ . }
It  is plausible, therefore, that it can be embedded into various
other spaces  containing $S^3$ factors, in particular into $AdS_5
\times T^{1,1}$  space related via AdS/CFT  to an $N=2$
superconformal theory  \kw.

Similarly,  analogs of $AdS_5$ two-spin solution of section 3
exist for a  more general  class of 5-d (or higher-dimensional)
metrics  with $SO(4)$ isometry, e.g.,
$ds^2 = - g(\r) dt^2 +d\r^2 + h(\r) d\Omega_3$.
As in the $AdS_5$ case, the two equal rotations  in $S^3$ allow
one to stabilize a circular string  at a fixed value of $\r=\r_0$
(stability under small fluctuations  will  depend on the
explicit form of $g(\r)$ and $h(\r)$  and on the value  of the
spin). In particular, such solution will exist  for an $AdS$
black hole metric.\foot{Single-spin  solutions in this and
similar ``non-conformal''   cases  were discussed in
\rf{\dev,\arm}.}

There are several directions in which  the
present work needs to be completed or extended. 
In \ftn\ we shall study  the  $S^5$ solution with  $J\not=0,
J'\not=0$ in detail,  identifying the range of its stability,
and 
computing the 1-loop  superstring sigma model correction to the
classical energy. One should then be able 
to   compare the string results  (e.g., for $J \sim  J'$) to the 
perturbative results for anomalous  dimensions of the
corresponding gauge theory operators. It would be very
interesting also to compute the leading-order  perturbative
contributions to the anomalous  dimensions of the operators
discussed in section 5. This applies, in particular, to the
operator \opr\  whose dimension should be possible to find using
the integrable-model connection  suggested in \miz.

\bigskip\bigskip
\noindent
{\bf Acknowledgements}

\noindent
We are grateful to G. Arutyunov, N. Beisert,
R. Metsaev, J. Minahan,   M. Staudacher and K. Zarembo for
useful discussions  and comments.
This work  was supported by the DOE grant DE-FG02-91ER40690.
The work of A.T.  was also supported in part by the  PPARC SPG 00613
and  INTAS  99-1590 grants and the Royal Society  Wolfson award.

\appendix{A}
{Stability analysis }
Here  we analyze the stability of the $AdS_5$ and $S^5$
two-spin solutions  discussed above
under small perturbations.

\subsec{\bf Stability of the two-spin $AdS_5$ solution}

 The equations of motion for fluctuations that follow
 from \lagri\ are
\eqn\eqmr{
\ddot{\td \r}-\tdr''-2(2+\k^2)\tdr -2\sqrt{2(1+\k^2)} \dot{\a}=0\ ,}
\eqn\eqma{
\ddot{\a}-\a''-2\sqrt{2+\k^2} \b' +2\sqrt{2(1+\k^2)}\dot{\tdr}=0\ ,}
\eqn\eqmb{
\ddot{\b}-\b''+4(1+\k^2)\b +2\sqrt{2+\k^2} \a' =0\ .}
To solve the system \eqmr--\eqmb\ of linear differential equations we
 expand the
fluctuations in Fourier series in $\s$
\eqn\you{
 \tdr = \sum_n\ \r_n(\tau )\ee^{i n \s}\ ,\ \ \ \a = \sum_n\ \a_n(\tau
)\ee^{i n \s}\ ,\ \ \
 \b = \sum_n\ \b_n(\tau )\ee^{i n \s}\ . }
Then we get the following system of equations for the $n$-th modes of the
fluctuations
\eqn\eqmrn{
\ddot{\r_n}+n^2\r_n -2(2+\k^2)\r_n -2\dot{\a_n}\sqrt{2(1+\k^2)}=0\ ,}
\eqn\eqman{
\ddot{\a_n}+n^2\a_n +2\dot{\r_n}\sqrt{2(1+\k^2)}-2i n\b_n\sqrt{2+\k^2} =0\
,}
\eqn\eqmbn{
\ddot{\b_n}+n^2\b_n+4(1+\k^2)\b_n +2in\a_n\sqrt{2+\k^2} =0\ .}
We start the analysis of the  system \eqmrn-\eqmbn\ by looking for
solutions of the form \eqn\oscil{ \r_n(\tau ) =  C^\r_n \ee^{i \omega_n
\tau}\ , \ \ \ \a_n(\tau ) =  C^\a_n \ee^{i \omega_n \tau}\ ,
\ \ \ \b_n(\tau ) =  C^\b_n \ee^{i \omega_n \tau}\ .}
The stability of the two-spin string solution requires the
frequencies $\omega_n$ to be real. Substituting \oscil\
 into \eqmrn,\eqman,\eqmbn\
 we get
\eqn\eqmrC{
[n^2 - 2(2+\k^2) -\omega_n^2]C^\r_n - 2i\omega_n\sqrt{2(1+\k^2)}C^\a_n =0\
,}
\eqn\eqmaC{
(n^2 - \omega_n^2)C^\a_n + 2 i\omega_n\sqrt{2(1+\k^2)}C^\r_n - 2i
n\sqrt{2+\k^2}C^\b_n =0\ ,}
\eqn\eqmaC{
[n^2 +4(1+\k^2) - \omega_n^2]C^\b_n + 2i n\sqrt{2+\k^2}C^\a_n =0\ .}
This is a system of linear equations for the coefficients $C^\r_n,\
C^\a_n,\ C^\b_n$ which can be written in the form $A_{ij} C^j  =0$; it has
nontrivial solutions only for such values of $\omega_n$ for which the
determinant of the matrix $A_{ij}$ vanishes. This gives the equation for
$\w_n$:
$$ f_n(z) = 0 \ , \ \ \ \ \ \ \ \   z\equiv\omega_n^2  \ , $$
\eqn\detA{
f_n(z) \equiv z^3-(8 + 10\kappa^2 + 3n^2)z^2
+( 16 + 40\kappa^2 + 24\kappa^4 + 8\k^2n^2 + 3n^4)z
 - n^2( n^2 -4)(n^2  -4 - 2\kappa^2)
\ . }
We need to find the values of $\k$ such that all roots of this cubic
equation are positive. It is not difficult to show that
the two extrema $z_-$ and $z_+$ \ ($z_- < z_+$) of $f_n(z)$ are positive,
 and
$f_n(z_-)>0$ and $f_n(z_+)<0$. Therefore, all roots are positive if
\eqn\roo{
f_n(0)= - n^2( n^2 -4)(n^2 - 4 - 2\kappa^2) \le 0\ .  }
We see that $f_0(0) = f_2(0) = 0$, $f_1(0)< 0$ and
$f_3(0)= -45(5-2\k^2)$. From the expression for $f_3(0)$
we conclude that the solution is stable only if
\eqn\vall{
\k^2\le {5\ov 2}\ .
}
We also see that for $n=0$ and $n=2$ the
determinant of the matrix  $A_{ij}$ in \detA\
vanishes at $\omega_n =0$. That means that for these
modes there is a solution of the form
\eqn\linear{
\r_n(\tau ) = \tilde{C}^\r_n \tau + C^\r_n\ ,\ \
\ \ \ \a_n(\tau ) = \tilde{C}^\a_n \tau + C^\a_n \ ,
\ \ \ \ \ \ \b_n(\tau ) = \tilde{C}^\b_n \tau + C^\b_n \ .}
Substituting \linear\ into the equations of motion \eqmrn, \eqman\ and
\eqmbn, one can easily find the corresponding solutions for $n=0$ and
$n=2$
\eqn\lino{
\r_0(\tau )=C^\r_0\ , \ \ \ \ \a_0(\tau )=-{2+\k^2\ov
\sqrt{2(1+\k^2)}}C^\r_0\ \tau +C^\a_0\ , \ \ \ \ \b_0(\tau )=0\
,} \eqn\linii{
\r_2(\tau )=C^\r_2\ ,\  \ \ \a_2(\tau )=-{\k^2\ov
\sqrt{2(1+\k^2)}}C^\r_0\ \tau  + C^\a_2\ , \ \
\b_2(\tau )= -{i\ov\sqrt{2+\k^2}}\a_2(\tau )
\ .}
Since the fluctuation $\td \r$ corresponding to
\lino,\linii\ does not depend on $\tau$,
 such
solutions do not lead to instability of
 the two-spin string solution.
In fact,  both solutions \lino\ and
\linii\  have simple interpretation.
The $n=0$ solution reflects the fact that the
 radius $\r_0$ (or $\k$)
of the two-spin string solution \same--\soll\
is a free parameter. This parameter can be
changed, and this leads to the existence of the zero mode in the spectrum
of fluctuations. The $n=2$ solution  appears
 because the
two-spin string solution was found only for equal frequencies.
We expect that the general solution
will  depend on the two independent   frequencies,
and, therefore, on the  two parameters;
the existence of the zero mode fluctuation with $n=2$ is related
to this (cf. \rety).

\subsec{\bf Stability of the two-spin $S^5$ solution}

 The analysis of stability in the $S^5$ case follows
 closely  the
procedure explained above in the $AdS_5$ case.
Here we shall consider explicitly
only the case of the  solution with $J=0$
and $\JJJ \leq \ha$ discussed in section 4.3.
 The equations of
motion for fluctuations that follow  from \lagrS\ are
\eqn\eqmgS{
\ddot{\td \g}-\td\g''-2\mu^2\td\g +2\mu\dot{\a}=0\ ,}
\eqn\eqmaS{
\ddot{\a}-\a''-2\sqrt{2}\b' -2\mu\dot{\td\g}=0\ ,}
\eqn\eqmbS{
\ddot{\b}-\b''+4\b +2\sqrt{2}\a' =0\ .}
Expanding the fluctuations $\td \g$, $\a$ and $\b$
in Fourier series in $\s$, and then  looking for solutions in
the form $\ee^{i\w_n\tau}$, we find the following equation for the
frequency spectrum ($\m^2 \equiv 2-\k^2$)
$$ f_n(z) = 0 \ , \ \ \ \ \ \ \ \   z\equiv\omega_n^2  \ , $$
\eqn\detAS{
f_n(z) \equiv z^3-(4 + 2\mu^2 + 3n^2)z^2
+( 8\mu^2 + 3n^4)z
 - n^2( n^2 -4)(n^2  - 2\mu^2)
\ . }
Just as in the $AdS_5$ case, the two extrema  $z_- < z_+$ of $f_n(z)$
are positive, with  $f_n(z_-)>0$ and $f_n(z_+)<0$. Therefore, all roots are
positive if
\eqn\rooS{
f_n(0)= - n^2( n^2 -4)(n^2 -  2\mu^2) \le 0\ .  }
Since $\mu^2 \leq 2$, the solution is stable only if $f_1(0) =
3(1- 2\mu^2) \le 0$, i.e. if
$${1\ov 2}\le \mu^2 \leq 2\ .$$

\appendix{B}
{Quadratic fermionic part of  the superstring  action}

The quadratic part of the \adss  Green-Schwarz superstring action \MT\
expanded near a particular bosonic string solution  can be found as
described, e.g., in \rf{\KT, \dgt}  and used in a similar context
in
 \fts.
Assuming the induced metric is flat, the relevant  part of the
fermionic action is
\eqn\fer{
L_F=
i (\eta^{ab }\delta^{IJ} -
\ep^{ab } s^{IJ} ) \bar \vt^I \vr_a D_b \vt^J   \ ,\ \ \ \ \ \
\vr_a \equiv \G_{A}  e^A_a \ , \ \ \  e^A_a= E^{A}_M \del_a X^M \ ,
}
where
 $I,J=1,2$,  $s^{IJ}=$diag(1,-1), and
  $\vr_a$ are projections of the 10-d Dirac
matrices.
Here $X^M$ are the  string coordinates
(given functions of $\ta$ and $\s$ for a particular classical
solution)
corresponding to the  \ads ($M=0,1,2,3,4$) and $S^5$
($M=5,6,7,8,9$)
factors.
The covariant derivative $D_a$
can be put into  the  following form
\eqn\form{
D_a\t^I   = (\delta^{IJ} {\rm D}_a
- { i \ov 2 } \epsilon^{IJ}  \G_* \vr_a ) \vt^J\ ,
\ \ \ \ \ \  \G_* \equiv i \G_{01234} \ , \ \
\G_*^2 =1 \ ,
 }  where
 \eqn\topp{ {\rm D}_a = \del_a
+\fourth    \omega^{AB}_a\Gamma_{AB} \ ,
\ \ \ \ \ \ \ \omega^{AB}_a \equiv   \del_a  X^M \omega^{AB}_M
\ , }
 and
 the ``mass term''  originates from the R-R  5-form
 coupling  \MT.

 \subsec{\bf $AdS_5$ case}

In the case of the $AdS_5$ two-spin solution \hop\
the  2-d projections  of $\G$-matrices that enter  the
fermionic action  are (the indices $A=0,1,2,3,4$  here will
label the $t,\r,\t,\p,\vp$ directions in the
tangent space):
\eqn\varr{
 \vr_0 = \k \cosh \r_0 \ \G_0 +  \w \sinh \r_0 \td \G_4 \ , \ \ \
 \ \ \ \vr_1 =
 \sinh \r_0 \ \G_2 \ ,  \ \ \ \ \ \
  \ \   \vr_{(a} \vr_{b)} = \sinh^2 \r_0
\ \eta_{ab} \ ,  }
\eqn\vraa{ \td \G_3 \equiv  \cos \s \ \G_3 - \sin \s \ \G_4 \ , \ \
\ \ \ \
 \td \G_4 \equiv  \cos\s\ \G_4   +    \sin\s \ \G_3  \  . }
The projected Lorentz connection $\omega^{AB}_a=
\del_a  X^M \omega^{AB}_M$
has the following components
$$\omega^{01}_0 = \k \sinh \r_0 \ , \ \ \ \ \ \
\omega^{31}_0 = \w \cosh \r_0\sin \s  \ , \ \ \
\omega^{41}_0 = \w \cosh \r_0\cos\s  \ , $$  \eqn\coom{
\omega^{32}_0 = -\w \cos \s  \ , \ \ \ \ \ \ \
\omega^{42}_0 = \w \sin\s \ ,  \ \ \ \ \ \ \
\omega^{21}_1 = \cosh \r_0  \  .  }
Then
\eqn\dew{
{\rm D}_0 = \del_0 + \ha  (\k \sinh\r_0  \G_0 +  \w \cosh \r_0 \td
\G_4) \G_1
+  \ha \w \G_2 \td \G_3 \ , \ \ \ \ \ \
{\rm D}_1 = \del_1 -  \ha \cosh \r_0 \ \G_1 \G_2 \ . }
After the $\s$-dependent
rotation of $\theta^I$  in the 34-plane,
$\theta^I\to \Psi^I = S^{-1} \theta^I, \
S= \exp ( {1\ov 2} \s \G_3 \G_4 ) $, we  find
$\td \G_{3,4} \to \G_{3,4}$ in \varr\ and \dew, at the expense
of
getting an additional constant $\G_3 \G_4$ term in D$_1$.
This eliminates the $\s$-dependence from the fermionic action.
Note that $\G_*$ is invariant under this rotation.

To interpret the resulting fermionic action as a collection of
massive 2-d fermions with standard kinetic terms it is useful to
make
as in \fts\ a further ``rotation'' (Lorentz boost) in the 04-plane,
with $S= \exp ( \ha \a \G_0 \G_4) $, where
$\cosh \a = \k \cosh \r_0$. Then  we get $\vr_a =
 \sinh\r_0 \ \tau_a,
\ \  \tau_a = ( \G_0,\G_2), \ \tau_{(a}\tau_{b)}=\eta_{ab}$.
There is  the corresponding change in D$_a$
while $\G_*$ remains  invariant.
 Fixing the kappa-symmetry gauge
  by  $\Psi^1=\Psi^2$ and rescaling the fermions by $\sinh\r_0$
    we  can interpret the resulting action
  \eqn\fuu{
L_F = 2i
 \left( \bar \Psi \tau^a {\rm D}_a
\Psi   +  i \bar \Psi  M  \Psi \right)
\ , \ \ \ \ \ \  }\eqn\masa{
M =  \ha \sinh \r_0 \ \ep^{ab} \tau_a  \G_*
\tau_b =   i m_{_{F}}  \G_{134} \ ,  \ \ \ \   \
  m_{_{F}}= \sinh \r_0 = {1\ov \sqrt 2}\k \
}
as describing
a collection of 2-d  massive Majorana fermions on
a flat 2-d background
coupled
to a constant non-abelian 2-d  gauge field (represented by
constant Lorentz connection  $\w^{AB}_a$ terms). Indeed,
in the representation for  $\G_A$ where
$\G_0$ and $\G_2$ are  2-d Dirac matrices times a unit $8\times 8$
matrix
we get as in \fts\   4+4 species of 2-d Majorana fermions  with
masses
$\pm m_{_{F}}$.

 We will not go into a detailed
analysis of this action here and just mention that, as expected
on the general grounds
 of conformal invariance of the \adss string action \MT,
the fermionic contribution to  the divergent part of the 1-loop
effective action (which is proportional to the sum of mass-squared
terms,
i.e. $ 8 \times { \k^2 \ov 2}$)
 cancels the logarithmic divergence
coming from the bosonic fluctuation action \laak\
(connection terms in both the bosonic and fermionic actions do not
contribute  to logarithmic divergences in 2 dimensions).

 \subsec{\bf $S^5$ case}

 In the case of the $S^5$ solution \ann\
 we shall label the tangent space coordinates
 by  $A=0,5,6,7,8,9$  corresponding to the
$t$ direction of $AdS_5$ and $\g,\vp_1,\psi,\vp_2,\vp_3$ directions
of $S^5$.
 Then  \eqn\varr{
 \vr_0 = \k  \G_0 +  \n \cos \g_0 \ \G_6  + \ww \sin\g_0 \ \td
 \G_8\ , \ \ \
 \ \ \ \vr_1 =
 \sin \g_0 \ \G_7 \ ,  \ \ \ \ \ \
  \ \   \vr_{(a} \vr_{b)} = \sin^2 \g_0
\ \eta_{ab} \ ,  }
\eqn\vrua{ \td \G_8 \equiv  \cos \s \ \G_8  +  \sin \s \ \G_9 \ , \
\ \ \ \ \
 \td \G_9 \equiv  \cos\s\ \G_9   -    \sin\s \ \G_8 \  . }
The projected Lorentz connection
has the following non-zero components
$$\omega^{65}_0 = -\n \sin\g_0  \ ,  \ \ \ \   \ \ \
\omega^{85}_0 = \ww\cos\g_0\   \cos \s  \ , \ \ \ \ \ \
\omega^{95}_0 = \ww \cos \g_0\ \sin\s  \ , $$
  \eqn\conm{\ \ \
\omega^{87}_0 = -\ww \sin \s  \ , \ \ \ \ \ \
\omega^{97}_0 = \ww \cos\s  \ , \ \ \ \ \ \
\omega^{75}_1 = \cos \g_0 \ .  }
As above, we first do local Lorentz rotation in the 89-plane to
eliminate
the $\s$-dependence; as a result,  $\td \G_{8,9} \to \G_{8,9}$.
Then (for generic $\n$ and $\g_0$)
we need to do two rotations  -- in the 68 and 06 planes --
to put $\vr_0$ into the form $\vr_0 = \sin \g_0 \ \G_0$.
After the  rotation in the 68-plane (under which $\G_*$ in \form\
is
invariant) we get $\vr_0 = \k \G_0 + a \G_6, \
a^2 = \nu^2 \cos^2 \g_0 + \ww^2 \sin^2 \g_0 = \n^2 + \sin^2 \g_0 $.
Under  the boost in the 06-plane $S= \exp( \ha \b \G_0 \G_6)$,
where $\cosh \b = { \k \ov \sin\g_0}$,  the expression for
$\G_*$ becomes
\eqn\bec{
\G_*'= S^{-1} \G_* S = i ( \cosh \b \ \G_0 - \sinh \b\ \G_6)
\G_{1234} \ . }
Then fixing the kappa-symmetry gauge
  by  $\Psi^1=\Psi^2$ and rescaling the fermions by $\sin\g_0$
    we  finish with the same action as
    in \fuu\ with $\tau_a= (\G_0,\G_7)$  and
\eqn\maso{
M =   \ha \sin \g_0 \ \ep^{ab} \tau_a  \G'_*
\tau_b =   i m_{_{F}}  \G_{07} \G_{12346} \ ,  \ \ \ \   \
 m_{_{F}}= \sin \g_0\ \sinh \b  = { 1 \ov \sqrt 2}
 \sqrt {\k^2 + \nu^2} \   .
}
The contribution to the  divergences is then proportional
 to
$8 \times \ha ( \k^2 + \n^2)= 4 \k^2 + 4 \n^2$
  which is indeed the same
 as coming from the bosonic sector (see also Appendix
C).

The same result is found
also in the special cases discussed in section 4.3.
When $\nu=0$  (see \thenn)  we have
 $ \vr_0 = \k  \G_0   +  \sin\g_0 \ \td \G_8, \ \
  \vr_1 = \sin \g_0 \ \G_7 \ ,\ \  \sin \g_0 = {\k \ov \sqrt 2}$.
 Here we need a boost in 08-plane  with parameter
 $\cosh  \b = \sqrt 2$ to get  $ \vr_a = \sin \g_0\  \tau_a$.
 Then $m_{_{F}}$ is the same as in \maso.
When  $\g_0 = {\pi \ov 2}$, \ $\ww^2 = \k^2 -1$ (see \taq)\foot{
 Note that for $\cos \g_0=0$ the connection \conm\ simplifies
  substantially.}
we get
  $ \vr_0 = \k  \G_0   +  \ww  \td \G_8, \ \ \vr_1 =  \G_7$
  and thus the required 08-boost parameter  has
   $\cosh  \b = \k$. That gives
 \eqn\gii{\g_0 = {\pi \ov 2}\ : \ \ \ \
 M=   i m_{_{F}}  \G_{07} \G_{12348}
 \ , \ \ \ \ \ \ \
 m_{_{F}}=  \sinh \b  =\sqrt {\k^2 -1 } \ . }
 Note that the fermionic and bosonic  masses are different,
 reflecting the absence of the 2-d supersymmetry.
 The fermionic contribution to the
 logarithmic divergence is
  proportional to
$8 \times  ( \k^2 -1 )= 4 \k^2 + 4 (\k^2 -2) $
 which indeed  cancels  the  contribution from the bosonic fluctuations:
 4  massive $AdS_5$ fields \adsd\ and  4 massive
 $S^5$ fields in \sph.

 \appendix{C}{ Bosonic fluctuation action  in conformal gauge }

In discussing fluctuation actions in the main part of the paper we used
the static gauge.  Let us note for completeness that  similar
conclusions can be   reached also if one uses
  the conformal gauge (see, e.g.,  \rf{\dgt,\fts}).
Let us recall that
 the general form of the quadratic fluctuation action
for a sigma model in the conformal gauge
written  in terms of tangent-space
fluctuations  ($
X^M \to  X^M + \z^M ,$
$\z^A = E^A_M(X)  \z^M$)  is
\eqn\ctoq{
I^{(2)}_{ B} = -{ 1 \ov 2 } \int d^2 \xi
\left( \sqrt {-g} g^{ab} {\rm D}_a \z^A {\rm D}_b \z^B\  \eta_{AB}
+ M_{AB} z^A z^B \right),
}
\eqn\ytr{
M_{AB} = - \sqrt{-g} g^{ab}  e_a^{\,C} e_b^{\,D} R_{ACBD}\,,
\ \ \ \ \
e_a^{\,A} \equiv \del_a  X^M E^A_M ( X) \ , \ \ \ \   g_{ab} =
e^A_a e^B_b
\eta_{AB} \ ,
}
where  D$_a \z^A = \del_a \z^A + \omega^{AB}_a (X)   \z^B$ with the
same
projected Lorentz connection as in \topp.
For
the \ads part   the curvature is
$ R_{ACBD }=-\eta_{AB }\eta_{CD}+\eta_{AD}\eta_{CB}$
while for the $S^5$ part it has the opposite sign,
$ R_{ACBD }=\d_{AB }\d_{CD}- \d_{AD}\d_{CB}$.
If the induced metric is flat (as in all examples discussed
in the present paper), the
 divergent part of the 1-loop action is determined  simply by the
 trace
 of $M_{AB}$ and should be the same as found in the static
 gauge, cf. \fts.

 Let list the expression for the  mass matrix in \ytr\
 for the solutions discussed above
 (the expressions for the corresponding connections
 can be found in Appendix B).
 In the $AdS_5$ case \hop\ one gets:
 $$ e^0_0 = \k \cosh \r_0, \ \ \ \ \ e^2_1 = \sinh\r_0, \  \ \ \ \
 e^2_0 = \w \sinh \r_0\ \cos\s, \   $$ $$
 e^3_0 = \w\sinh \r_0\ \sin \s\ , \ \ \ \ \ g_{ab} = \sinh^2\r_0\
 \eta_{ab}\ , $$
 so that  $\eta_{AB} M^{AB} = M^2_{AdS_5} $,
 \eqn\iuy{
 M^2_{AdS_5}=4 \eta^{ab}  e_a^{\,C} e_b^{\,D} \eta_{CD}= 4 (
 \k^2 \cosh ^2 \r_0  + \sinh^2\r_0 - \w^2 \sinh^2 \r_0)  =
 4 \k^2 \ . }
 This  gives the same contribution
 to divergences as  coming   from the
 fermions (cf. \masa).

 For the $S^5$ solution \ann\
  one has  both the $AdS_5$  and $S^5$
 fluctuations, and
 $$ e^0_0 = \k, \ \ \ \ \ \ e^6_0 = \n \cos \g_0, \ \ \ \ \ \
 e^7_1 = \sin\g_0, \  $$ $$   e^8_0 = \ww \sin \g_0\ \cos\s, \
\ \ \ \  \  e^9_0 = \ww \sin \g_0\ \sin \s\ , \  \ \ \ \ \
 \ g_{ab} = \sin^2\g_0\ \eta_{ab} \ ,     $$
 so that here
 \eqn\dio{
 \eta_{AB} M^{AB}    = M^2_{AdS_5} +  M^2_{S^5}\ , \ \ \ \ \ \ \ \ \
  M^2_{AdS_5}=4 \eta^{ab}  e_a^{\,C} e_b^{\,D} \eta_{CD}=
 4 \k^2 \ , }
 \eqn\kol{   M^2_{S_5}=
 - 4 \eta^{ab}  e_a^{\,C} e_b^{\,D} \delta_{CD}=
 -4 ( \sin^2 \g_0 -\n^2 \cos^2 \g_0 -  \ww^2 \sin^2  \g_0 )
 = 4 \n^2  \ .}
 This is  in agreement with  the static gauge result \lagrS\
 for
 $\n=0$ and is the same as the divergent contribution coming
 from the fermionic sector with mass matrix  \maso.

 In the special case  $\g_0={\pi\ov 2},\ \ww^2= \k^2 -1$
 we get instead
 \eqn\jopw{
   M^2_{S_5}  = 4 (\k^2 -2)   \ , }
which is also in  agreement with the
static gauge result \sph\ and again  cancels the divergences
 coming from the fermionic sector \sph.

\appendix{D}{Dynkin labels and Young tableau labels
of $SU(4)$ irreps }

In this Appendix we recall (see, e.g., \jon)
how the Dynkin labels
of representations of the algebra
 $so(6)$ or, equivalently, of $su(4)$,
  are
expressed in terms of
 the Young labels (numbers of boxes in  raws
 of a  Young tableau).    Recall that the generators
$J_1=J_{12},\ J_2=J_{34},\ J_3=J_{56}$ form a
basis of Cartan
generators of $so(6)$. The simple roots can be chosen as
\eqn\rootso{
\a_1^{so(6)} = e_1-e_2 =\{1,-1,0\}\ ,\ \ \ \a_2^{so(6)} = e_2-e_3
=\{0,1,-1\}\ ,\ \ \ \a_3^{so(6)}= e_2+e_3 = \{0,1,1\}\ .}
Since $\a_2^{so(6)}\cdot \a_1^{so(6)} = \a_1^{so(6)}\cdot
\a_3^{so(6)} = -1$ and $(\a_a^{so(6)})^2 = 2$
 the root
system is equivalent to the $su(4)$ root system with the
following identification of the simple roots
\eqn\rootsu{
\a_1^{su(4)}=\a_2^{so(6)}\ , \ \ \ \a_2^{su(4)}=\a_1^{so(6)}\ ,\
\ \ \ \a_3^{su(4)}=\a_3^{so(6)}\ , }
i.e. the  two algebras are isomorphic.

An  irreducible representation of $su(4)$
 can be labelled by the
eigenvalues of the Cartan generators $J_i$ on the highest weight
vector:
\eqn\ji{
J_i|j_1,j_2,j_3\ra  =j_i |j_1,j_2,j_3\ra\ , \ \ \ \ \ \ \ \
 \ j_1\geq
j_2\geq j_3\geq 0\ ,}
where $j_i$ is the number of boxes in the $i$-th row of the
Young tableau associated with the representation which
can then be  denoted
as $(j_1,j_2,j_3)$. The same representation can be also labeled
by the Dynkin labels $d_a$ that are related to the Young labels
$j_i$ as follows: \eqn\jidy{
j= \sum_{a=1}^3 d_a \lambda_a\ ,\ \ \ \ \ \ \ \ \ \ \ \
j\equiv \{j_1,j_2,j_3\}\
,}
where $\lambda_a$ are the  fundamental weights defined by
\eqn\fundwe{
{2\lambda_a\cdot \a_b\ov\a_b^2} = \lambda_a\cdot \a_b =\d_{ab}\
.}
By using \rootso\ and \rootsu, we can easily solve \fundwe\ to
get
\eqn\fundwei{
\l_1 =\{\ha,\ha,-\ha\}\ ,\ \ \ \ \ \  \ \l_2 =\{1,0,0\}\ ,\ \ \ \ \
\ \
\l_3 =\{\ha,\ha,\ha\}\ .}
Then from \jidy\ and \fundwei\ we find
\eqn\jidyi{
j_1 = \ha( d_1 + 2d_2+d_3)\ , \ \ \ \ \ \ \ j_2 = \ha( d_1 +d_3)\ ,
\ \ \ \ \ \ \    j_3 = \ha( -d_1 + d_3)\ .}
Solving the system, we get the relation between
the Dynkin labels
and the Young labels
\eqn\dyji{
d_1 = j_2-j_3\ , \ \ \ \ \ \ \
\ d_2 = j_1-j_2\ , \ \ \ \ \  \ \ d_3 = j_2+j_3 \
.}
The Dynkin labels have to be non-negative. The  representation
associated  with the Dynkin labels $d_i$ is denoted as
$[d_1,d_2,d_3]$.

Coming back to the string solution \takk, we see that if $J\geq J'$
then the representation is $(J,J',J')$ or $[0, J-J',2J']$. If
$J'\geq J$ the representation is $(J',J',J)$ or $[J'-J,0,J'+J]$.
Note that in the last case we also have to rearrange the Cartan
generators in such an order that $j_1\geq j_2\geq j_3\geq 0$.

\vfill\eject
\listrefs

\end